\documentclass[10pt,twocolumn,final,journal]{IEEEtran}

\usepackage{bigstrut,multirow,array,psfig,epsfig,subfigure,amssymb,amsmath,color,url,xr}
\usepackage[ruled,linesnumbered]{algorithm2e}

\begin{document}
\pagenumbering{arabic}

%Title option #1
\title{Multiple Target Tracking with RF Sensor Networks}
%Title option #2
%\title{Device-Free Tracking of Multiple Targets with RF Sensor Networks}
%Title option #3
%\title{Tracking of Multiple Targets\\with RF Sensor Networks}

\author{Maurizio Bocca\thanks{Maurizio Bocca and Neal Patwari are with the Department of Electrical and Computer Engineering, University of Utah, Salt Lake City, UT, USA. E-mail: maurizio.bocca@utah.edu, npatwari@ece.utah.edu}, Ossi Kaltiokallio\thanks{Ossi Kaltiokallio is with the Department of Automation and Systems Technology, Aalto University, Helsinki, Finland. E-mail: ossi.kaltiokallio@aalto.fi}, Neal Patwari \IEEEmembership{Member,~IEEE} and Suresh Venkatasubramanian\thanks{Suresh Venkatasubramanian is with the School of Computing, University of Utah, Salt Lake City, UT, USA. E-mail: suresh@cs.utah.edu}}
%\markboth{IEEE Transactions on Mobile Computing}{Bocca et al.: Tracking of Multiple Targets with RF Sensor Networks}
\maketitle

\begin{abstract}
RF sensor networks are wireless networks that can localize and track people (or targets) without needing them to carry or wear any electronic device. They use the change in the received signal strength (RSS) of the links due to the movements of people to infer their locations. In this paper, we consider real-time multiple target tracking with RF sensor networks. We perform radio tomographic imaging (RTI), which generates images of the change in the propagation field, as if they were frames of a video.  Our RTI method uses RSS measurements on multiple frequency channels on each link, combining them with a fade level-based weighted average. We describe methods to adapt machine vision methods to the peculiarities of RTI to enable real time multiple target tracking. Several tests are performed in an open environment, a one-bedroom apartment, and a cluttered office environment. The results demonstrate that the system is capable of accurately tracking in real-time up to 4 targets in cluttered indoor environments, even when their trajectories intersect multiple times, without mis-estimating the number of targets found in the monitored area. The highest average tracking error measured in the tests is $0.45$ m with two targets, $0.46$ m with three targets, and $0.55$ m with four targets.

\end{abstract}

%\category{C.4}{Performance of Systems}{Design Studies}
%\category{J.3}{Life and Medical Sciences}{Health}
%\terms{Algorithms, Measurement}

%Sections
\section{Introduction} \label{sec:introduction}
\PARstart{R}{adio} frequency (RF) sensor networks are wireless networks that monitor the changes in the received signal strength (RSS) of the links in order to localize and track people, without requiring them to carry or wear any radio device. In these systems, the RSS measurements can be processed to form images of the changes in the propagation field of the monitored area in presence of moving people and objects - a process named radio tomographic imaging (RTI) \cite{wilson09a}. Unlike previous works \cite{wilson09a,kanso09b,chen11sequential,kaltiokallio2011} which have focused on single target localization and tracking, this work aims at contributing to the growing field of device-free localization (DFL) through RF sensor networks \cite{wilson09a,patwari10c} by presenting a system capable of accurately tracking in real-time multiple people (or \emph{targets}) moving in real-world indoor environments where low-power transceivers are deployed.

The potential applications of RF sensor networks are many, including smart buildings and perimeter surveillance, ambient assisted living and residential monitoring \cite{kaltiokallio2012b}, breathing detection \cite{breathing_Patwari}, and security and rescue operations. Compared to other sensing technologies applied in indoor DFL, such as infrared, ultrasonic range finders \cite{whitehouse_2012}, ultra-wideband (UWB) radios and video cameras, RF sensor networks provide several advantages: they work in the dark and can penetrate smoke and walls; they are less invasive in domestic environments than video camera networks; they are significantly less expensive than UWB transceivers; their installation and maintenance time is minimal.

People moving in an area where wireless transceivers are deployed affect the propagation of the radio signals by shadowing, reflecting, diffracting or scattering a subset of their multipath components \cite{patwari10c,hashemi94,ghaddar2004human,wilson10see}. In open and uncluttered environments, where line-of-sight (LoS) communication among the transceivers is predominant, a person obstructing a link line will generally cause attenuation of the radio signal. This phenomenon has been succesfully applied for DFL in several works \cite{wilson09a,kanso09b,chen11sequential,kaltiokallio2011}. However, in cluttered environments, where multipath propagation is predominant, the change in RSS due to the presence of a human body becomes more unpredictable. As the link line is obstructed, the RSS can also remain constant or increase \cite{wilson10see}. In addition, due to the multipath propagation of the radio signals, people can affect the RSS also when located far away from the link line \cite{kaltiokallio2012a}.

Most of the research presented so far in the area of DFL with RF sensor networks has considered the situation in which only one target had to be located and tracked \cite{wilson09a,kanso09b,chen11sequential,kaltiokallio2011}. However, the real-world scenarios in which these systems are to be used often require the localization and tracking of multiple targets. Moreover, the DFL system should be able \emph{a)} to perform the localization and tracking tasks in real-time even when the targets trajectories intersect, and \emph{b)} to correctly estimate the number or targets to be tracked as people enter and exit the monitored area. In this work, we tackle all of these challenges. In forming the RTI images, we apply a novel approach that weights the RSS measurements based on the fade level \cite{wilson11fade} of the frequency channels on which the RSS was measured. We adapt machine vision methods to RTI and use the new methods to process the RTI images in real-time to detect and track the blobs corresponding to real targets. Multiple target tracking with RF sensor networks is made more challenging in that in RTI the targets have to be modeled as points, consisting only of position, velocity and acceleration, neither having physical length or other features such as shape, color, or size. 
Although the blob size and shape in RTI are loosely related to a person's size, the blob dramatically changes depending on a person's position in the monitored area and on the positions of other people in the area. Thus, the noise in the blob shape overwhelms any attempt to determine a person's features. We invite the reader to view an RTI video at \cite{MTT_videos} as an example of the capabilities and limitations of the imaging modality.
In addition, due to measurement noise and the simultaneous presence of multiple targets, objects and obstructions in the monitored area, spurious blobs (not corresponding to real targets) can appear in the image. For the same reasons, blobs corresponding to real targets can temporarily disappear from the image. These factors increase the difficulty of multiple target tracking, especially with intersecting trajectories.

We evalute the performance of the multi-target tracking system in three different indoor environments, \emph{i.e.} an open environment with no obstructions nor objects, a one-bedroom apartment with internal walls, furniture and various objects, and a heavily cluttered office environment. The few multi-target tracking methods developed for RSS-based DFL are either non-real time \cite{wilson11fade,Zhang_2009}, track only two or three people \cite{wilson11fade,Zhang_2009,Zhang_2011,MTT_tracking_Nannuru2012}, assume that the number of targets is fixed and known a priori \cite{wilson11fade,MTT_tracking_Nannuru2012,MTT_tracking_Nannuru2011}, or do not attempt to track targets with intersecting trajectories \cite{wilson11fade,Zhang_2009,Zhang_2011,MTT_tracking_Nannuru2012,MTT_tracking_Nannuru2011}. In an experimental test with four targets having separated trajectories, the method in \cite{Thouin_2011} achieves as low as $0.63$ m average accuracy, consuming $7.6$ seconds to process an RTI image. Our method achieves $0.55$ m average accuracy in an experimental test with four targets having intersecting trajectories, consuming $13.3$ milliseconds to process an RTI image. To the best of our knowledge, we are the first to demonstrate that RF sensor networks can be used in real-world indoor environments to accurately track in real-time up to four targets even when their trajectories intersect multiple times. The highest average tracking error measured in the tests is $0.45$ m with two targets, $0.46$ m with three targets, and $0.55$ m with four targets. Videos showing the performance of the multi-target tracking system during the tests described in this paper can be found at \cite{MTT_videos}.

The remainder of the paper is organized as follows. We present the methods used to form radio tomographic images and to track multiple targets in Section \ref{sec:multichannel_RTI} and \ref{sec:MTT}, respectively. The experiments carried out are described in Section \ref{sec:experimental_setup}, and the results are listed and discussed in Section \ref{sec:tests_results}. The related work is described in Section \ref{sec:related_work}. Conclusions are drawn in Section \ref{sec:paper_conclusion}.

\section{Multi-channel RTI} \label{sec:multichannel_RTI}
In this section, we describe how the RSS measurements collected on multiple channels are processed in real-time to form radio tomographic images. We deploy $R$ sensors at positions $\{\mathbf{z}_{r}\}_{r=1,...,R}$. At time instant $k$, we measure the RSS $r_{l,c}(k)$ in dBm of link $l$ on channel $c \in {\{1,...,K\}}$. Our objective is to estimate the change in the propagation field of the monitored area, $\mathbf{x}$, from the RSS measurements collected on all the links of the network.

\subsection{Fade level}

In obstructed environments, radio signals propagate from the transmitter to the receiver via multiple paths. At the receiver, a phasor sum of the waves impinging on the antenna determine the RSS. Depending on the relative phase of each wave, the waves may add constructively or destructively. As a wave's phase is a function of the center frequency and path length, the RSS is a function of the center frequency and position of the communicating devices, an effect called multipath fading.

The relation between  steady-state, narrow-band fading and the temporal fading statistics of  the RSS due to human movement has been described in \cite{wilson11fade}, where the authors define the concept of \emph{fade level}, a continuum between two extremes: \emph{deep fade} and \emph{anti-fade}.  For a link in a deep fade, when the link line is obstructed, the RSS, on average, increases. In addition, deep fade links show changes of the RSS  even when the person is at some positions far away from the link line. On the  contrary, for a link in an anti-fade, when the link line is obstructed, the RSS, on average, decreases. Moreover, anti-fade links show changes of the  RSS only when the person is in the close proximity of the link line. Due to the limited size and predictable shape of their sensitivity area, the anti-fade links are the most informative for DFL. In \cite{kaltiokallio2012a,kaltiokallio2012b}, the channels used to form RTI images were sorted based on their fade  level, and the RSS measurements of most anti-fade channels of each link  were selected. In this work, we use channel diversity and propose a  novel approach to consider the RSS measurements of all the channels and  weight them based on their fade level.

During an initial calibration period performed in static conditions, \emph{i.e.}, when the monitored area is empty, we measure the average RSS of each link $l$ on each different frequency channel $c$, $\bar{r}_{l,c}$, which can be modeled as:
\begin{equation} \label{E:average_RSS}
    \bar{r}_{l,c} = \tilde{P}_{c}+G_{l,c}, 
\end{equation}
where $\tilde{P}_{c}$ is the actual transmit power, in dBm, on channel $c$ and $G_{l,c} < 0$ is the path gain, in dBm, of link $l$ on channel $c$. For the wireless sensors used in the experiments (see Section \ref{sec:hardware}), the actual transmit power $\tilde{P}_{c}$ is different from the nominal one and, most importantly for estimating the fade level, is a function of the frequency channel $c$, in part due to the difficulty in antenna impedance matching across a wide frequency band \cite{tidongleantenna}.

The fade level of channel $c$ for link $l$ is estimated as the difference between the path gain measured for channel $c$ and the minimum of the path gains measured for the used frequency channels:
\begin{equation} \label{E:fade_level}
    F_{l,c} = G_{l,c}-\min_{c}G_{l,c}.
\end{equation}
Thus, for the same link $l$, channel $c_1$ is in a deeper fade than channel $c_2$ if $F_{l,c1} < F_{l,c2}$. Note that $F_{l,c} \ge 0$, and that $F_{l,c} = 0$ for one channel $c$ on each link.

\subsection{Measurement model}

The RSS measurements collected on different channels are weighted based on the fade level defined in (\ref{E:fade_level}), bearing in mind that anti-fade channels provide measurements more informative for localization. The weighted average change in RSS of link  $l$ at time $k$ is computed as:
\begin{equation}\label{E:RSS_change}
    y_{l}(k) =  \frac{1}{\sum_{c\in{K}}F_{l,c}} {\sum_{c\in{K}}F_{l,c} \cdot \lvert \Delta r_{l,c}(k) \rvert},
\end{equation}
where $\Delta r_{l,c}(k)$ is the difference between the RSS of link $l$ on channel $c$ measured at time $k$, $ r_{l,c}(k)$, and the reference RSS, $\bar{r}_{l,c}$.

\subsection{Image estimation}

The change in RSS is assumed to be a spatial integral of the propagation  field of the monitored area. When the propagation field is discretized, some voxels affect the RSS of a specific link, whereas others do not. Thus, the change in RSS of each link is assumed to be a linear combination of the change caused by each voxel:
\begin{equation}\label{E:linear_combination}
    y_{l}(k) =  \sum_{j=1}^N w_{lj}x_{j}+n_{l},
\end{equation}
where $x_{j}$ is the change in RSS caused by voxel $j$, $w_{lj}$ the weight of voxel $j$ for link $l$, $N$ the number of voxels in the discretized image and $n_l$ the noise of link $l$. The weight $w$ indicates how each voxel of the image affects each link. For this, we use an ellipse model \cite{wilson09a,wilson10see,kaltiokallio2012a} in which the transmitter and receiver are located at the foci of the ellipse. 
According to this model, a voxel $j$ at position $\mathbf{v}_{j}$ that is located within the ellipse of link $l$ has its weight $w_{lj}$ set to a constant, which is inversely proportional to the area of the ellipse. Otherwise its weight is set to zero, as follows:
\begin{equation}\label{E:weight model}
    w_{lj} = \begin{cases}
                       \frac{1}{A_l} & \text{if } d_{lj}^{tx}+d_{lj}^{rx}<d_{l}+\lambda\\
                       0 & \text{otherwise}
                                  \end{cases},
\end{equation}
where $d_{lj}^{tx} = \| \mathbf{z}_{l_{tx}}-\mathbf{v}_{j} \|$ and $d_{lj}^{rx} = \| \mathbf{z}_{l_{rx}}-\mathbf{v}_{j} \|$, $d_l = \| \mathbf{z}_{l_{tx}}-\mathbf{z}_{l_{rx}} \|$, $A_l$ is the area of the ellipse, and $\lambda$ is its excess path length, \emph{i.e.}, the parameter defining the width of the ellipse.

When all the links of the wireless network are considered, the change in the propagation field of the monitored area is:
\begin{equation}\label{E:linear_formulation}
    \mathbf{y} =  \mathbf{W}\mathbf{x}+\mathbf{n},
\end{equation}
in which $\mathbf{y}$ and $\mathbf{n}$ are $M \times 1$ vectors representing the weighted RSS change and noise of the $M$ wireless links, and $\mathbf{x}$ is the $N \times 1$ change in the propagation field to be estimated, where each element $x_j$ represents change due to the presence of a person in voxel $j$. The linear model for the  change in the propagation field is based on the correlated shadowing  models in \cite{patwari08b,agrawal09} and on the work in \cite{wilson09a}.

Since estimating the image vector $\mathbf{x}$ from the links' measurements $\mathbf{y}$ is an ill-posed inverse problem, regularization is required. In this work, we use a regularized least-squares approach \cite{patwari08b,zhao11noise}:
\begin{equation}\label{E:linear_transformation}
    \hat{\mathbf{x}} =  \mathbf{\Pi}\mathbf{y},
\end{equation}
where:
\begin{equation}\label{E:tikhonov}
    {\mathbf{\Pi}} = {(\mathbf{W}^T\mathbf{W}+\mathbf{C}_{x}^{-1}\sigma_{N}^{2})}^{-1}\mathbf{W}^T,
\end{equation}
in which $\sigma_{N}^{2}$ is the regularization parameter. The \emph{a priori} covariance matrix $\mathbf{C}_{x}$ is calculated by using an exponential spatial decay:
\begin{equation}\label{E:cov_matrix}
    [\mathbf{C}_{x}]_{ji}=\sigma_{x}^{2}e^{-\| \mathbf{v}_{j}-\mathbf{v}_{i} \| /\delta_{c}},
\end{equation}
where $\sigma_{x}^{2}$ is the variance of voxel measurements, and $\delta_{c}$ is the voxels' correlation distance. The linear transformation $\mathbf{\Pi}$ is computed only once before real-time operation. The calculation of $\hat{\mathbf{x}}$ in (\ref{E:linear_transformation}) requires $MN$ operations and can be performed in real-time.

\subsection{Image denoising}

An RTI image representing the situation in which $n$ targets are located in different regions of the monitored area should ideally show $n$ \emph{blobs}, \emph{i.e.}, regions in which the voxels have an intensity much higher than the intensity of the surrounding voxels. However, due to noise in the RSS measurements  and the simultaneous presence of multiple individuals and obstructions, the RTI images are often noisy, showing multiple spurious blobs of small  size which do not correspond to actual targets. For this reasons, a Gaussian filter is applied on the RTI image. This operation has the effect of reducing the image's high spatial frequency components, filtering out small spurious blobs while simultaneously keeping those larger blobs corresponding to actual targets \cite{davies2012computer}.

The filtering is obtained by convolving the RTI image with an isotropic Gaussian kernel:
\begin{equation}\label{E:gaussian_kernel}
    {\mathbf{G}(x,y)} = \frac{1}{{2}{\pi}{\sigma_{G}^{2}}}e^{-{\frac{x^{2}+y^{2}}{{2}{\sigma_{G}^{2}}}}},
\end{equation}
where $\sigma_{G} = 1$ m is the standard deviation of the Gaussian kernel which indicates how much the image is filtered (or \emph{blurred}). Since the estimated RTI image $\hat{\mathbf{x}}$ is stored as a set of discrete voxels, we need to produce a discrete approximation of the Gaussian kernel $\mathbf{G}$ to be able to perform the convolution. Thus, the kernel is truncated after $\left \lfloor {r_G}/{p}+{0.5} \right \rfloor$ voxels, where $r_G = 0.75$ m is the radius of the kernel. The low-pass filtered RTI image $\hat{\mathbf{x}}_{G}$ is then calculated as:
\begin{equation}\label{E:denoised_image}
    {\hat{\mathbf{x}}_{G}} = {\hat{\mathbf{x}}} \ast {\mathbf{G}},
\end{equation}
where $\ast$ represents  the convolution operator. As a result, each voxel of the filtered RTI image is a weighted average of the voxel's neighborhood, with the central pixels weighted more than the peripheral ones. The values of $\sigma_G$ and $r_G$ are chosen so as to fit the size of the blobs corresponding to the targets.

\section{Multiple Target Tracking} \label{sec:MTT}
The methods described in the previous section form radio tomographic images in real-time, \emph{i.e.}, pictures of the change in the propagation field of the monitored area caused by the presence and movements of the people found in it. These images can be considered as \emph{frames} of a video showing the movements of multiple targets. In this section, we describe the methods used to process the RTI images and track multiple targets in real-time. Our objective is to correctly detect the entrance and exit of the targets and estimate their position $\mathbf{v}_t$ by assigning to each of them a voxel of the estimated RTI image.

\subsection{Thresholding}
\label{sec:thresholding}

After denoising the image, an additional filter is required in order to reduce the size of the set of voxels that go through the clustering process (see Section \ref{sec:clustering}) and to preserve only those in the regions occupied by the targets. To this purpose, a dynamic threshold $T_{t}$ is set as follows.

During the calibration period, \emph{i.e.}, when the monitored area is empty, we calculate the average maximum intensity of the formed RTI images, $\bar{I}_e$. When the monitored area is empty, the threshold $T_{t}$ is set to $2\bar{I}_e$ to filter the voxels having very low intensity. On the other hand, when targets are being tracked, we derive the minimum intensity, ${I}_{min}$, of the targets $t$ in the set of estimated targets $\mathcal{T} = \{t_1, \ldots, t_{|\mathcal{T}|}\}$ as:
\begin{equation}\label{E:minimum_intensity}
    {I}_{min} = {\min_{t \in \mathcal{T}} [{\hat{\mathbf{x}}_{G}}]_{t}}
\end{equation}
We then low-pass filter ${I}_{min}$:
\begin{equation}\label{E:low_pass_filt_imin}
    {I_{f}(k)} = \alpha_{f} I_{f}(k-1) + ({1}-{\alpha}_{f}) I_{min}(k),
\end{equation}
where $\alpha_{f} = 0.9$. Thus, the threshold $T_{t}$ is set as follows:
\begin{equation}\label{E:threshold_targets_mask}
    T_{t}(k) = \begin{cases}
                   {\beta} I_{f}(k) & \text{if $|{T(k)}|$ $>$ $0$}\\
                   2\bar{I}_e & \text{otherwise}
               \end{cases},
\end{equation}
where $|{T(k)}|$ is the number of targets found in the monitored area at time instant $k$ and $\beta < 1$. In the experiments, we could not observe any significant performance change for values of $\beta$ in the range $[0.75,0.9]$.
The on-line updating of $T_{t}$ ensures that the voxels surrounding the tracks are not filtered and will go through the clustering process.

A vector mask $\mathbf{M_t}$ of size $N \times 1$ is created as follows:
\begin{equation}\label{E:targets_mask}
    [\mathbf{M_t}]_{n} = \begin{cases}
                   1 & \text{if } [{\hat{\mathbf{x}}_{G}}]_{n} > {{T}_{t}}\\
                   0 & \text{otherwise}
               \end{cases}.
\end{equation}
The denoised and filtered RTI image $\hat{\mathbf{x}}_{f}$ is finally calculated as the element-wise product of the denoised image $\hat{\mathbf{x}}_{G}$ and $\mathbf{M_t}$:
\begin{equation}\label{E:hadamard_product}
    {\hat{\mathbf{x}}_{f}} = {\hat{\mathbf{x}}_{G}} \wedge {\mathbf{M_t}}.
\end{equation}

\subsection{Clustering}
\label{sec:clustering}

In the clustering phase, voxels are assigned to each blob found in the image. Since we do not make any a priori assumption on the number of targets to be tracked (new targets can enter the monitored area at any time as tracked targets can leave it, and spurious blobs can also appear), clustering algorithms, such as \emph{k-means} \cite{kmeans}, for which the number of clusters must be known a priori can not be applied. For this reason, we use an hierarchical agglomerative clustering (HAC) algorithm \cite{hastie2009elements}.

We define $\mathcal{V}$ as the set of voxels that are not filtered in the thresholding phase (see Section \ref{sec:thresholding}), \emph{i.e.}, $\mathcal{V} = \{ j: [{\hat{\mathbf{x}}_{f}}]_{j} > T_t\}$. Each voxel $j \in \mathcal{V}$ has coordinates $\mathbf{v}_j = [x_j,y_j]$ in the XY plane.
In the HAC algorithm, each voxel is initially considered as an independent cluster. At each iteration, the two closest clusters are merged. The distance between two clusters, $S_a \subset \mathcal{V}$ and $S_b \subset \mathcal{V}$, is measured with the average linkage distance ${\bar{d}}$, \emph{i.e.}, the average of the Euclidean distances between all the voxels assigned to the two clusters:
\begin{equation} \label{E:ALD_equation}
    {\bar{d}}(S_{a},S_{b}) = \frac{\sum_{{j_a} \in {{S}_{a}}} \sum_{{j_b} \in {{S}_{b}}} {\| \mathbf{v}_{j_a}-\mathbf{v}_{j_b} \|}}{|{{S}_{a}}| |{{S}_{b}}|}.
\end{equation}
The iterations terminate when the minimum of the average linkage distances among the clusters is larger than a threshold $T_c$, which determines the average size of the formed clusters, and ultimately their number, (\emph{i.e.}, several small clusters for low values of $T_c$, few larger clusters for high values of $T_c$).

We normalize the intensity of the image in the range $[0,1]$, so that the normalized intensity $\tilde{I}_j$ of each voxel $j \in \mathcal{V}$:
\begin{equation} \label{E:normalized_intensity}
    \tilde{I}_j = \frac{[{\hat{\mathbf{x}}_{f}}]_{j}}{\max_{j \in \mathcal{V}} [{\hat{\mathbf{x}}_{f}}]_{j}}.
\end{equation}
For each formed cluster $S_i$, the voxel ${h}_i \in S_i$ having the maximum normalized intensity is selected as the cluster head:
\begin{equation} \label{E:cluster_head_selection}
    {h}_i = \arg\max_{j \in S_i} \tilde{I}_j.
\end{equation}
We define the set of \emph{original cluster heads} $\mathcal{H}$ as the set of voxels ${h}_i$ for all $i$.

\subsection{Cluster heads selection}
\label{sec:cluster_heads_selection}

Due to the 3D shape of the blobs found in the RTI image (typically round in open environments, having more distorted shapes in obstructed areas and when targets trajectories intersect), the HAC algorithm can form several cluster heads. In this case, we want to decrease the number of elements in $\mathcal{H}$ in order to reduce the complexity of the observations-targets association problem (see Section \ref{sec:targets_tracking}), and simultaneously keep, for every occupied region, only the cluster heads with the highest intensities, which are more likely to correspond to the real targets found in the monitored area. As a result of the selection process, a new set $\mathcal{H}_I \subseteq \mathcal{H}$ of cluster heads having higher intensities is formed.

In each indoor environment where the sensors are deployed, we define $\mathcal{R}_e \subset \mathcal{V}$ as the set of voxels included in the entrance/exit region, \emph{i.e.}, the region the targets must go through in order to enter and exit the monitored area. This region can be limited to a specific part of the monitored area (as in the apartment described in Section \ref{sec:apartment_environment}), or can cover the entire region along the perimeter of the monitored area (as in the open and office environments in Section \ref{sec:open_environment} and \ref{sec:office_environment}). All the cluster heads that are located within $\mathcal{R}_e$ are included in $\mathcal{H}_I$, regardless of their intensity.

The remaining cluster heads are selected based on their proximity to the targets being tracked and their intensity. As first step, gating is applied on the cluster heads in $\mathcal{H}$. At time $k$, for each cluster head $h$ at position $\mathbf{v}_{h}$, we define $\chi_{h}$ as:
\begin{equation}\label{E:gating_condition}
    \chi_{h} = \{ t: \| \mathbf{v}_{h} - \hat{\mathbf{v}}_{t} \| < {r_t}\},
\end{equation}
where $\hat{\mathbf{v}}_{t}$ is the position of the cluster head associated to target $t$ at time $k-1$.
The parameter $r_t$ represents the radius of the gating area centered at $\hat{\mathbf{v}}_{t}$. Initially, $r_t = r$ (see Table \ref{T:MTTParameters}). The value of $r_t$ is modified whenever the trajectory of target $t$ intersects the trajectory of another track (see Section \ref{sec:targets_intersection}). The gating area has to accommodate for the motion variance of the targets, indicating how fast the targets can move, and the typical noise of RTI images, (\emph{i.e.}, spurious and disappearing blobs, and blobs merging and splitting in case of intersecting trajectories).

Based on the results of the gating process, a cluster head $h$ is included in $\mathcal{H}_I$ if \emph{a}) $\| \chi_{h} \| \ge 1$, \emph{i.e.}, if there is at least one target $t$ whose gating area includes $h$, and \emph{b}) the normalized intensity of $h$ is larger than a threshold $T_h$, defined as:
\begin{equation}\label{E:min_intensity_gating}
    {T}_{h} = \rho \min_{t \in \chi_{h}} \tilde{I}_t,
\end{equation}
where $\rho = 0.8$.

\subsection{Target tracking}

\subsubsection{Tracks confirmation and deletion}
\label{sec:tracksconfdel}

The selected cluster heads in $\mathcal{H}_I$ are considered for updating the existing tracks and for potentially initiating new tracks. In machine vision, track confirmation and deletion is usually determined by rules \cite{blackman99}. In our case, the rules have to deal with RTI images that show new blobs when people enter the monitored area and stop showing blobs when people exit the monitored area. RTI images can show spurious blobs not corresponding to people, while blobs corresponding to real targets can temporarily disappear, as it is shown in the videos in \cite{MTT_videos}. Due to this, a trade-off exists between the reactivity of the system to the entrance and exit of targets and the sensitivity to the noise in the images.

At each frame, the data association methods presented in Section \ref{sec:targets_tracking} assign some of the cluster heads in $\mathcal{H}_I$ to the already existing tracks. Of the cluster heads which are not assigned, only those that are located in $\mathcal{R}_e$ are considered as new \emph{candidate tracks}, \emph{i.e.}, tracks potentially corresponding to new targets, whereas the ones located outside of $\mathcal{R}_e$ are considered as noise and are discarded. A candidate track becomes a \emph{confirmed track} only if it has been assigned a cluster head (see Section \ref{sec:targets_tracking}) at least $n_{app}$ times in the last $m$ frames ($n_{app} \le m$). The value of $n_{app}$ makes the system more or less reactive to the entrance of new targets. By using this rule, the system confirms the entrance of a new target only after having multiple confirmations of its presence. This introduces a small latency between when a new target enters the monitored area and the moment the system acknowledges it, as shown in Figure \ref{F:cardinality_figure}. However, the rule makes the system more robust to the appearances of spurious blobs in $\mathcal{R}_e$. Since the DFL system used in the experiments produces approximately $10$ RTI images (or frames) per second, we set $m=10$ so to consider the appearances of the tracks in a one second time interval before confirming their presence.

On the other hand, a track (whether confirmed or candidate) is deleted after it has not been assigned a cluster head in the last $n_{del}$ consecutive frames. Also in this case, the DFL system deletes the existence of the tracks with a small delay compared to reality. However, the rule prevents the system from incorrectly deleting tracks that have not been associated a cluster head for a few consecutive frames due to noise in the RTI image.

\subsubsection{Target tracking}
\label{sec:targets_tracking}

\begin{figure}[t!]
  \begin{center}
    \includegraphics[width=\columnwidth]{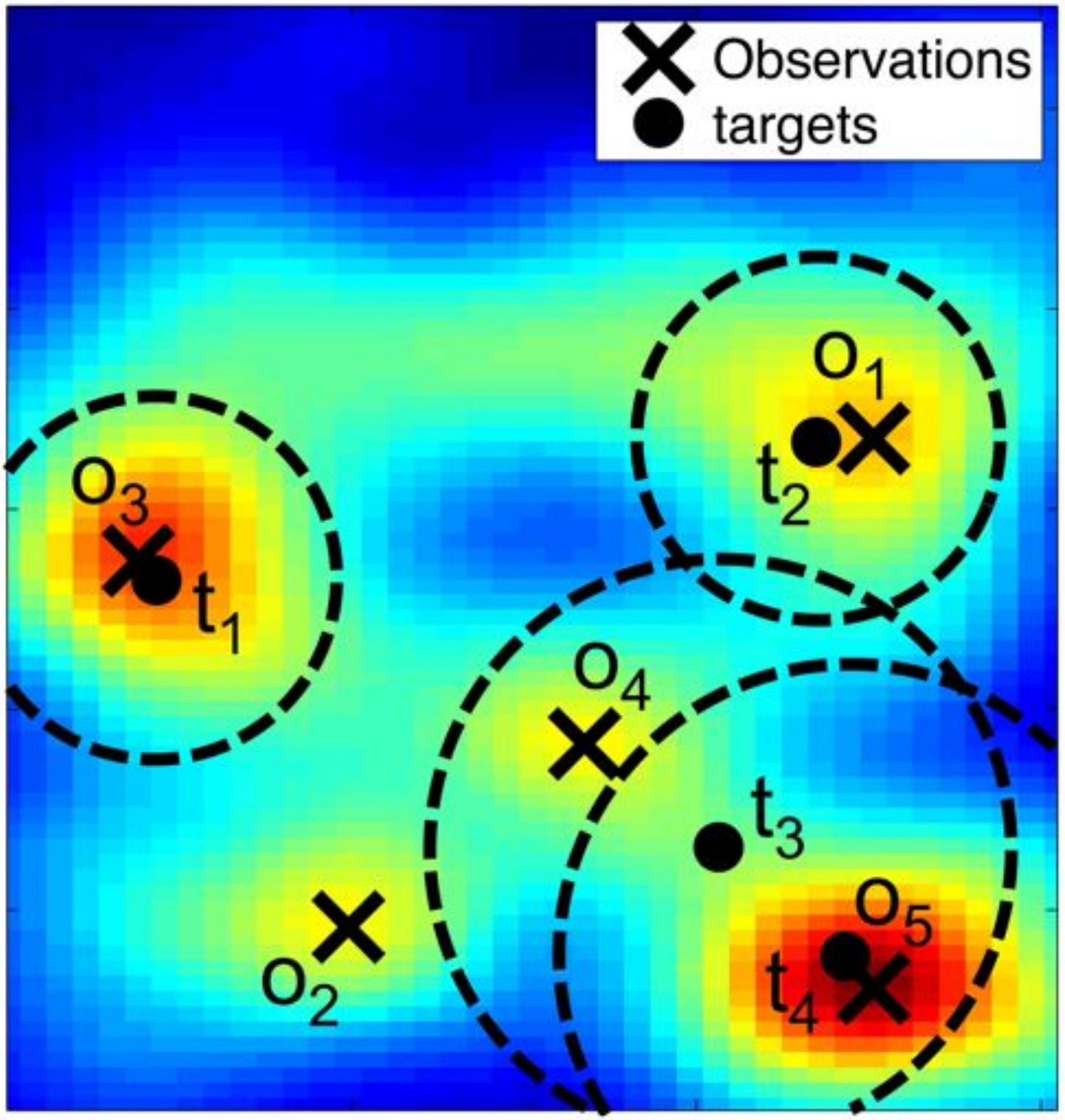}
    \caption{DAP in RTI. The dashed circles represent the gating areas centered at the tracks. In this case, the observation-tracks assignments are: $o_1-t_2$, $o_3-t_1$, $o_4-t_3$ and $o_5-t_4$. Observation $o_2$ is not assigned and is considered for starting a new track. The radius of the gating areas centered at $t_3$ and $t_4$ is larger as these two targets are intersecting.}
    \label{F:DAP_figure}
  \end{center}
\end{figure}

The problem of tracking multiple targets can be formulated as a data assignment problem (DAP), in which at each frame a set of new observations has to be assigned to the set of existing targets, as shown in Figure \ref{F:DAP_figure}. In our DFL system, at each new formed RTI image the set of selected cluster heads $\mathcal{H}_I = \{h_1, \ldots, h_{|\mathcal{H}_I|}\}$ has to be assigned to the set of estimated targets $\mathcal{T} = \{t_1, \ldots, t_{|\mathcal{T}|}\}$. The solution of this problem consists in finding the optimal permutation $\phi$ of the set $\mathcal{H}_I$, where the permutation matrix $\Theta$ is defined as:
\begin{equation}\label{E:permutation_matrix}
    \Theta_{h,t} = \begin{cases}
                       1 & \text{if } t=\pi(h)\\
                       0 & \text{otherwise}
                   \end{cases}.
\end{equation}
Based on the outcome of the gating process (see Section \ref{sec:cluster_heads_selection}), an association matrix $\mathbf{\Omega}$ is defined as follows:
\begin{equation}\label{E:assoc_matrix}
    [\mathbf{\Omega}]_{h,t} = \begin{cases}
        \| \mathbf{v}_{h} - \hat{\mathbf{v}}_{t} \| & \text{if } t \in {\chi}_{h}\\
        \infty & \text{otherwise}
    \end{cases}.
\end{equation}
The elements set to $\infty$ represent unfeasible observation-target assignments. When $\mathbf{\Omega}$ is interpreted as a cost matrix, the DAP becomes the problem of selecting  observation-target assignments that minimize the total cost.

\subsubsection{Prior work}
The DAP in multiple target tracking has received significant attention from  the research community, and several methods, including particle filters  (PF, \cite{Ristic_PF}), probabilistic and joint probabilistic data association methods (PDA and JPDA, \cite{JPDA_Bar_Shalom}) and multiple hypothesis tracking (MHT, \cite{ReidMHT}), have been proposed and analyzed \cite{MTT_taxonomy}. However, these methods present some limitations that make them not suitable to the requirements of our DFL system, which are \emph{a}) no a priori assumptions on the number of targets and \emph{b}) real-timeliness.

The PDA and JPDA approaches assume that the number of targets is known and constant. Moreover, the PF and MHT approaches are methods that struggle to meet strict real-time requirements: the PF needs a high number of particles to obtain accurate tracking, at the expense of a high computational complexity and long processing time \cite{wilson11fade,MTT_tracking_Nannuru2012}. Moreover, whenever the number of targets to be tracked increases, the PF needs a higher number of particles to maintain the same accuracy, making the processing time even longer. In MHT, each  feasible target-observation association is considered as an hypothesis  with a specific probability. At each new RTI image, each hypothesis is  expanded into a set of new hypothesis having specific probabilities, so  that a tree of hypothesis is incrementally generated. Each hypothesis  is expressed as a permutation matrix, and the method keeps a probability  distribution over the space of all permutation matrices. After several frames have been analyzed, the tracks having the highest probabilities  are selected. However, the number of hypothesis grows exponentially with the number of targets in $\mathcal{T}$, making this method computationally intensive \cite{Blackman_MHT}. Furthermore, the MHT approach is a \emph{batch} method, which postpones the DAP solution until more clear information is available. On the contrary, we aim at using a  recursive method that estimates the targets positions at time $k$ based on the observations available at time $k$ and the estimates of the targets positions at time $k-1$.

\subsubsection{Nearest neighbor methods}
In this work, two computationally efficient versions of the nearest neighbor (NN) approach, namely the global nearest neighbor (GNN) and the greedy (or suboptimal) nearest neighbor (SNN), are tailored for the characteristics of RTI. Despite its simplicity, the NN approach has demonstrated high tracking performance in scenarios characterized by noisy measurements, such as radar and sonar applications \cite{Leung_GNN,Konstantinova_GNN}. In GNN and SNN, the tracks are updated at each frame. A track can be  associated to only one observation, and in turn one observation can be  associated only to one track.

The GNN  method applies the Hungarian algorithm \cite{Kuhn1955,Munkres_rectangular}, which is capable of finding the optimal solution in a polynomial time $O(n^3)$, in which $n = \min (|\mathcal{T}|,|\mathcal{H}_I|)$. The SNN method applies a greedy approach to the selection of the observation-target assignments, which in $O(n)$ time is not guaranteed to find the optimal solution but requires fewer computations than the GNN method (especially when the number of feasible assignments is large). For the greedy SNN method, the upper bound of the total cost associated to the selection of the observation-target assignments is equal to twice the optimal  cost guaranteed by the GNN method. The observations that are not assigned to any target are considered for starting new tracks (see Section \ref{sec:tracksconfdel}). On the other hand, the tracks that are not assigned an observation are still predicted by using the Kalman filter (KF).

\subsection{Kalman filter tracking}
\label{sec:KF_tracking}

Whenever a candidate track $t$ is confirmed, a track-specific KF \cite{kalman1960,Welch95anintroduction} is initialized and recursively applied to track its movements. The system runs the KFs in parallel. Each KF estimates the new state, \emph{i.e.}, position, of a track by taking into consideration its previous state and the new observation, \emph{i.e.}, cluster head, associated to it (see Section \ref{sec:targets_tracking}). We assume that the targets to be tracked move as a Brownian process and that the measurement noise is Gaussian. The KF smoothes the trajectories of the targets and reduces their sudden changes of direction. At this purpose, it is particularly useful in the case of noisy RTI images.

\subsection{Handling intersecting trajectories}
\label{sec:targets_intersection}

In this work, we tackle the problem of tracking targets having intersecting trajectories. In RTI, this situation manifests itself in two (or more) blobs slowly merging into a single one and then splitting again after some frames.
In cluttered indoor environments, when two targets approach each other, the formed RTI images become very noisy due to the unpredictable overlap of the multiple propagation paths modified by each target. For this reason, whenever the distance between a target $t$ and any other target drops below $T_i$, $r_t$ in (\ref{E:gating_condition}) is doubled. With this adjustment, the motion variance of the targets is increased. Since in intersecting situations the merging-splitting blobs can disappear for several consecutive frames, increasing the radius of the gating area avoids losing track of targets.

\section{Experimental Setup} \label{sec:experimental_setup}
\begin{table}[t!]
    \caption{Image reconstruction parameters} %title of the table
        \centering
        \footnotesize
        \begin{tabular}{c c c} %centered columns (3 columns)
        \hline\hline\          %insert double horizontal line
        Parameter & Value & Description \\
        \hline  %insert single horizontal line
        $p$          & 0.1524 & Pixel width [m]                 \\ %body of the table
        $\lambda$    & 0.02   & Ellipse excess path length[m]   \\ %body of the table
        $\sigma_{x}$ & 0.2236 & voxels standard deviation [dB]  \\ %body of the table
        $\sigma_{N}$ & 1      & noise standard deviation [dB]   \\ %body of the table
        $\delta_{c}$ & 3      & correlation coefficient         \\ %body of the table
        \hline %inserts single line
        \end{tabular}
        \label{T:ImageParameters}
\end{table}

\begin{table}[t!]
    \caption{Multiple target tracking parameters (default values)} %title of the table
        \centering
        \footnotesize
        \begin{tabular}{c c c} %centered columns (3 columns)
        \hline\hline\          %insert double horizontal line
        Parameter & Value & Description \\
        \hline  %insert single horizontal line
        $\beta$      & 0.80   & Parameter used to set $T_t$ in (\ref{E:threshold_targets_mask})\\
        $T_e$        & 0.01   & Empty area intensity threshold \\
        $T_c$        & 1.25   & Voxels clustering threshold [m]               \\
        $T_i$        & 2      & Intersecting trajectories threshold [m]   \\
        $r$          & 2      & Radius of the gating area [m]             \\
        $r_G$        & 0.75   & Radius of the Gaussian kernel [m]         \\
        $\sigma_{G}$ & 1      & Std. deviation of the Gaussian kernel [m] \\
        \hline %inserts single line
        \end{tabular}
        \label{T:MTTParameters}
\end{table}

In this section, we describe the hardware and communication protocol used in the experiments and the environments in which we carried out the tests. The values of the image reconstruction parameters and of the multiple target tracking methods are listed in Table \ref{T:ImageParameters} and \ref{T:MTTParameters}, respectively.

\begin{figure*}[t!]
  \begin{center}
    \mbox{
      \subfigure[\quad open environment setup]{\includegraphics[width=2.1in,height=2.3in]{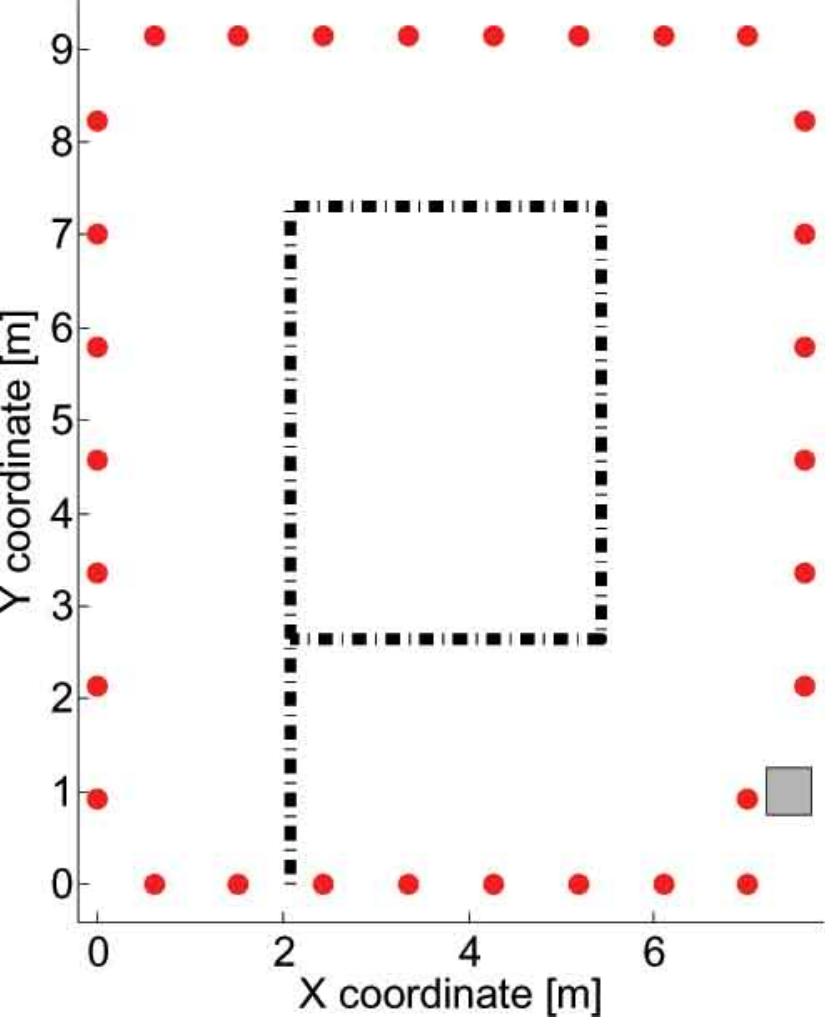}} \quad
      \subfigure[\quad apartment setup]{\includegraphics[width=2.1in,height=2.3in]{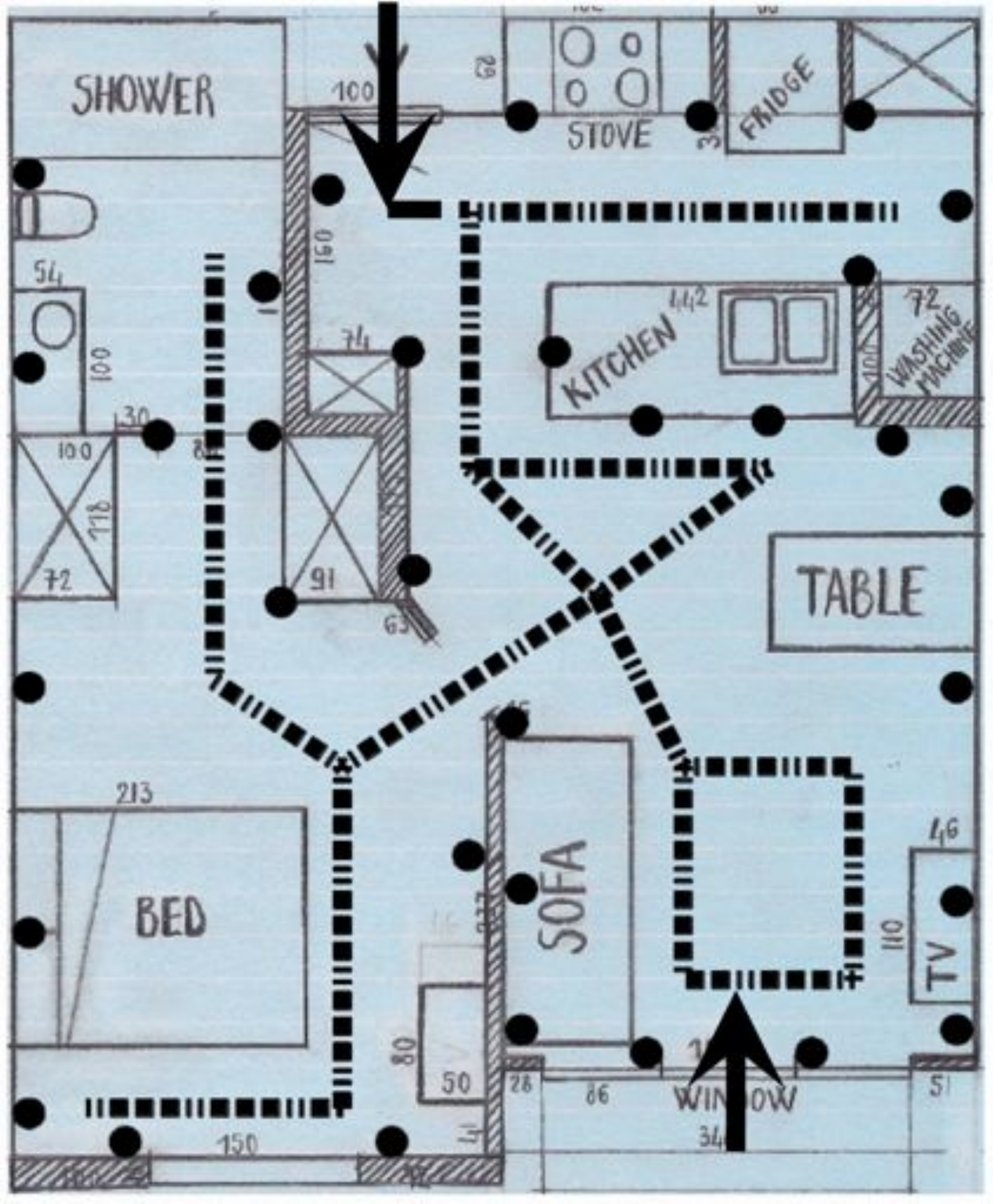}} \quad
      \subfigure[\quad office environment setup]{\includegraphics[width=1.7in,height=2.3in]{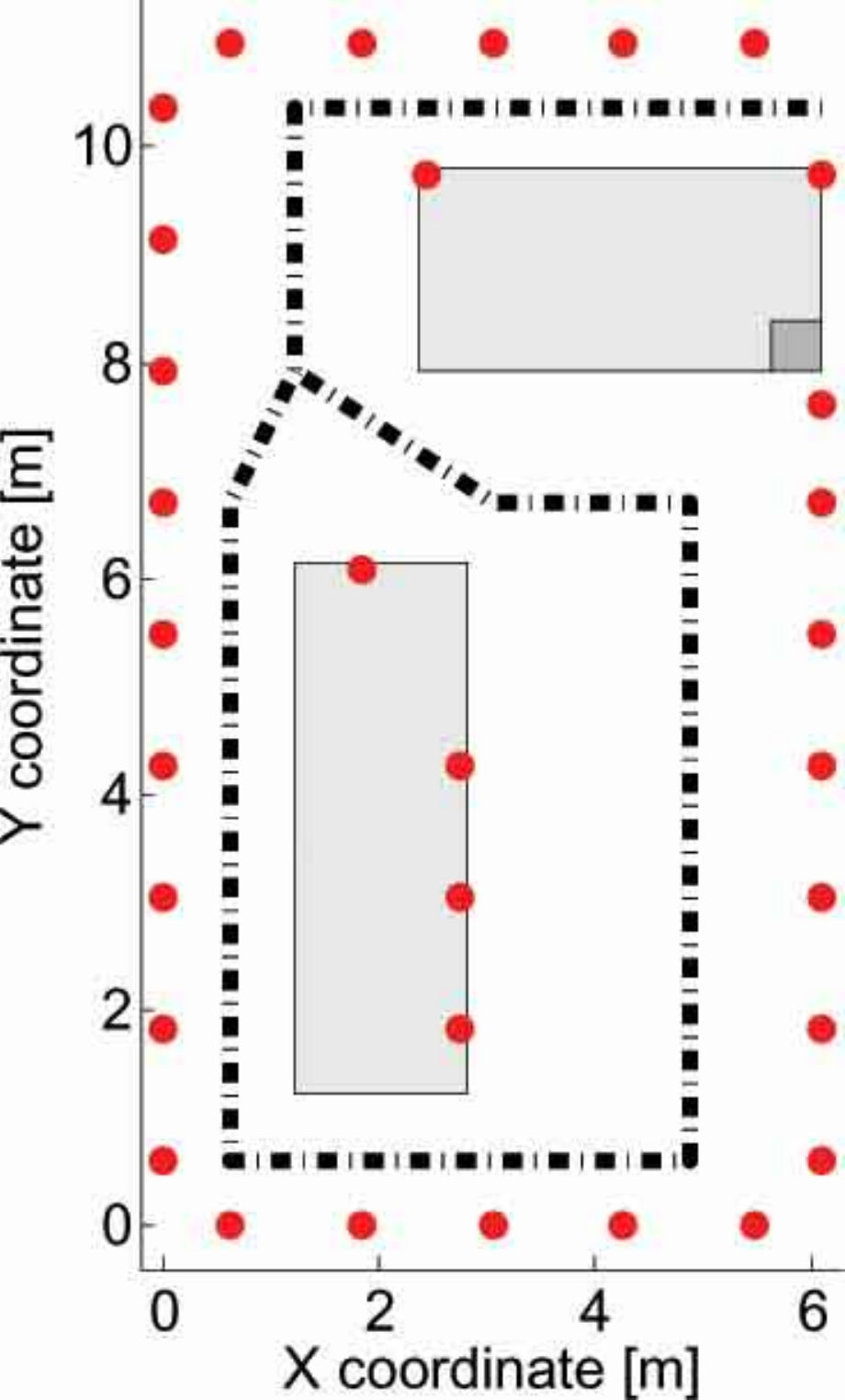}}
      }
    \mbox{
      \subfigure[\quad apartment environment]{\includegraphics[width=3.1in,height=2.0in]{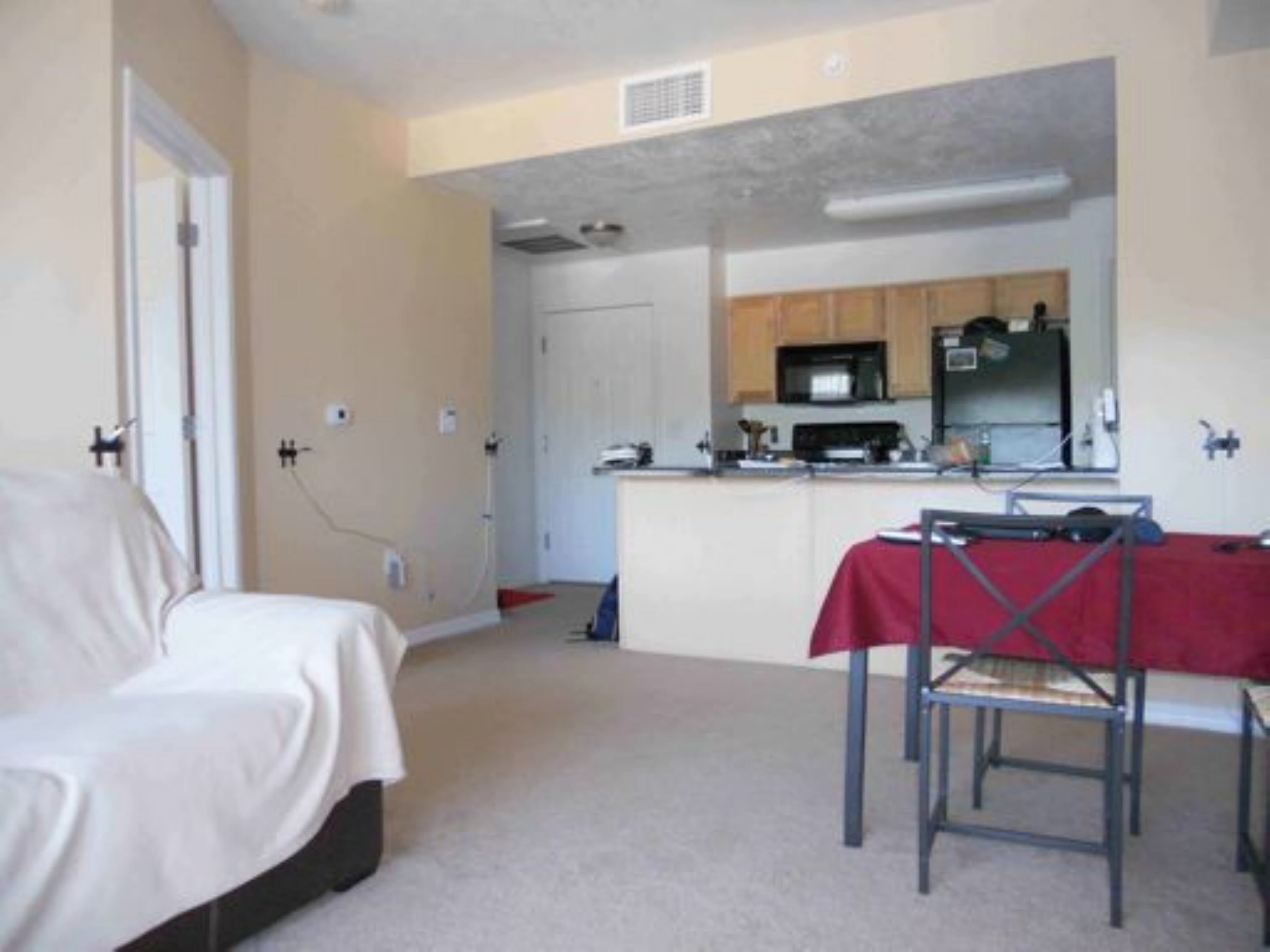}} \quad
      \subfigure[\quad office environment]{\includegraphics[width=3.1in,height=2.0in]{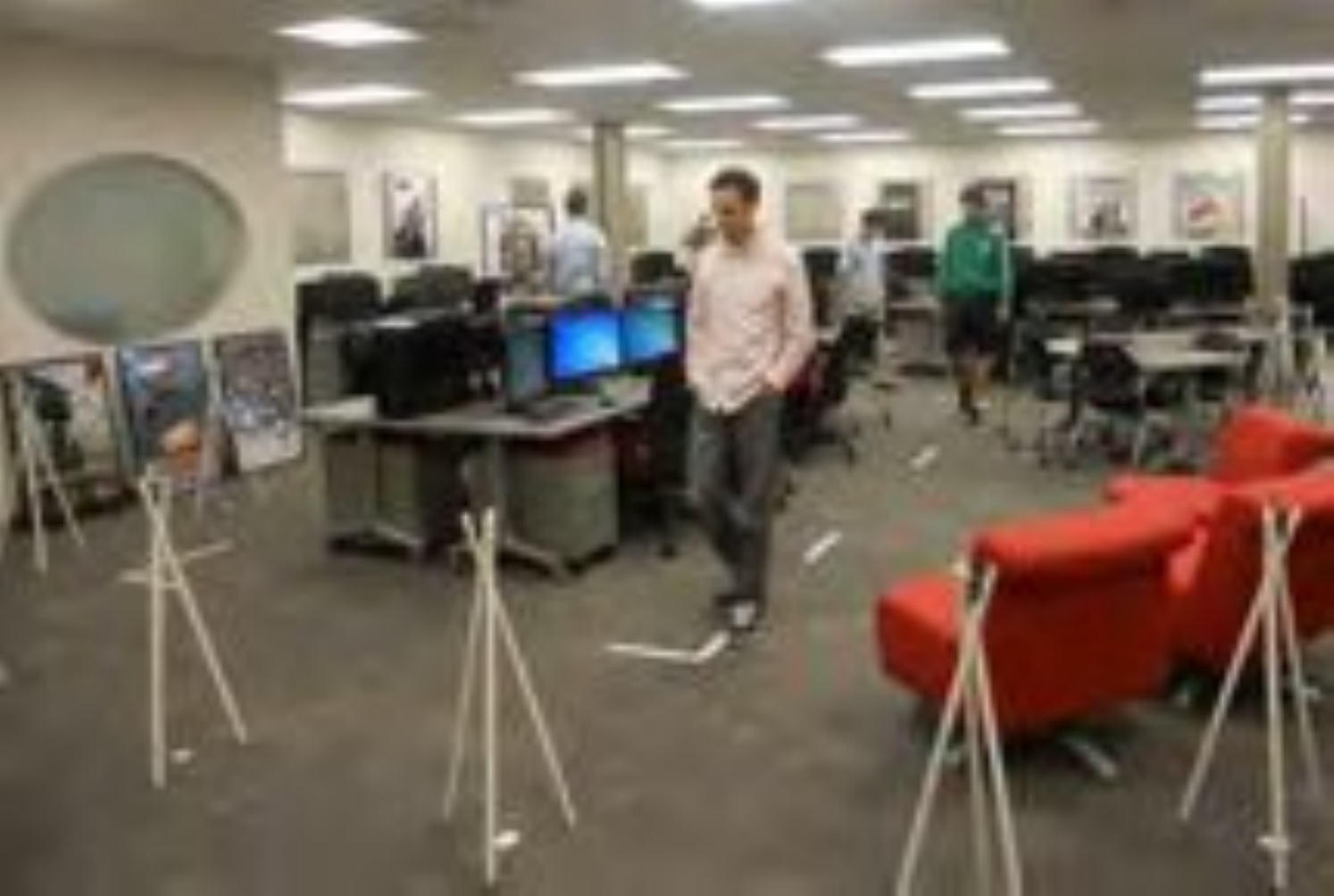}} \quad
      }
    \caption{In (a)-(c), the setups in the three indoor environments used for the experiments. The circles represent the sensors, while the dashed lines represent the paths followed by the people moving in the monitored area. The light gray rectangles in (c) are the rows of desks found in the office environment. The dark gray squares in (a) and (c) represent a concrete pillar. In (d), an image of the apartment. In (e), an image of the office environment.}
    \label{F:tests_setup}
  \end{center}
\end{figure*}

\subsection{Hardware and communication protocol}
\label{sec:hardware}

The sensors used in all the experiments are TI CC2531 USB dongle nodes, transmitting at their maximum nominal power, \emph{i.e.} 4.5 dBm \cite{tidongle}. For multi-channel communication, the sensors run a multi-channel token passing protocol, \emph{multi-Spin}, introduced in \cite{kaltiokallio2012a}. In it, the sensors transmit in TDMA fashion based on their ID number. Each transmitted packet contains the ID number of the transmitting node and the most recent RSS measurements of the packets received from the other sensors. At the end of each communication cycle, the sensors switch synchronously to the next frequency channel found in a list pre-defined by the user. On average, the time interval between two consecutive transmissions is 2.9 ms. A sink node that overhears all the packets transmitted by the nodes stores the RSS measurements for processing.

The CC2531 nodes transmit in the $2.4$ GHz ISM band in one of $16$ selectable frequency channels, which are $5$ MHz apart, as specified by the IEEE 802.15.4 standard \cite{802_15_4}.  The carrier frequency (in MHz) of channel $c$ is:
\begin{equation}\label{E:normalized_freq}
    f_{c} =  2405+5 \cdot (c-11),\quad c \in [11,26].
\end{equation}

\subsection{Test environments}
\label{sec:tests_description}

This section describes the three different indoor environments where the experiments were carried out. In all the deployment environments, multiple 802.11 b/g networks create interference in the $2.4$ GHz band \cite{srinivasan2006}.

\subsubsection{Open environment}
\label{sec:open_environment}

In an open indoor environment, \emph{i.e.} where no obstructions or objects are present, $30$ sensors are deployed along the perimeter of a $70 m^2$ area, as shown in Figure \ref{F:tests_setup}(a). The sensors are placed on podiums at a height of one meter from the floor. They transmit on $5$ different channels, \emph{i.e.} $c \in \{11,15,18,22,26\}$. During the tests, two people are instructed to walk at constant speed along a pre-defined rectangular path. In the first test, one person enters the monitored area and walks along the path, followed by the second person after few seconds. The two people walk along the path in the same direction, \emph{i.e.}, one behind the other. In the second test, the two people enter the monitored area few seconds one after the other, but this time the second person walks along the path in the opposite direction to the first person, so that the trajectories of the two people intersect. In this way, we are able to measure the accuracy of the proposed multiple target tracking system also when the targets converge to and then diverge from the same point in the monitored area. In this environment, we assume that people could enter and exit the monitored area at any point along the perimeter.

\subsubsection{Apartment}
\label{sec:apartment_environment}

We deploy $33$ sensors in a one bedroom, $58 m^2$ ($7$x$8.25$ m) apartment shown in Figures \ref{F:tests_setup}(b) and \ref{F:tests_setup}(d). The sensors are attached to the walls and furniture, at approximately one meter from the floor. They transmit on $4$ different channels, \emph{i.e.} $c \in \{15,20,25,26\}$. To reduce the noise of the RSS measurements \cite{kaltiokallio2012b}, the nodes are attached so to keep the antenna at least $5$ centimeters away from the wall. In the tests, two people walk along pre-defined paths. The trajectories intersect multiple times. We assume that the apartment has two specific entrance regions, located at the main door and at the sliding glass-door separating the balcony from the living room.

\begin{center}
\begin{table*}[t!]
    \caption{Processing time for the GNN, SNN, and MHT methods} %title of the table
        \centering
        \footnotesize
        \begin{tabular}{c c c | c c | c c | c c}
        \hline\hline %insert double horizontal line
          &  &  & \multicolumn{2}{c|}{GNN method} & \multicolumn{2}{c|}{SNN method} & \multicolumn{2}{c}{MHT method} \\
        \hline
        Targets & Environment & Intersections (\#) & $\max{[T_p]}$ [ms]& $E[T_p]$ [ms] & $\max{[T_p]}$ [ms] & $E[T_p]$ [ms] & $\max{[T_p]}$ [ms] & $E[T_p]$ [ms]\\
        \hline
        2 & open        & NO            & 15.4  &7.3          & 14.5  &7.0          &   16.8  &   7.3   \\ %body of the table
        2 & open        & YES (9)       & 13.5  &7.2          & 12.3  &7.0          &  622.5  & 134.6   \\ %body of the table
        \hline %insert single horizontal line
        2 & apartment   & YES (4)       & 19.4  &6.4          & 18.2  &5.9          &  712.9  & 121.1   \\ %body of the table
        2 & apartment   & YES (5)       & 19.3  &6.5          & 11.8  &5.8          &  537.7  & 109.2   \\ %body of the table
        \hline %insert single horizontal line
        2 & office      & NO            & 18.6  &6.8          & 12.2  &6.2          &   12.9  &   5.9   \\ %body of the table
        2 & office      & YES (5)       & 25.3  &7.4          & 14.2  &6.9          &  419.4  &  53.1   \\ %body of the table
        3 & office      & NO            & 26.1  &9.3          & 21.8  &8.8          &  231.6  &  28.0   \\ %body of the table
        3 & office      & YES (6)       & 30.3  &10.4         & 19.2  &9.2          & 1844.2  & 194.7   \\ %body of the table
        4 & office      & NO            & 35.6  &11.6         & 34.9  &11.4         &  321.3  &  34.6   \\ %body of the table
        4 & office      & YES (9)       & 43.4  &13.3         & 33.5  &11.5         & 4632.1  & 244.7   \\ %body of the table
        \hline %inserts single line
        \end{tabular}
        \label{T:proc_time_results}
\end{table*}
\end{center}

\subsubsection{Office environment}
\label{sec:office_environment}

In a typical office environment, shown in Figures \ref{F:tests_setup}(c) and \ref{F:tests_setup}(e), containing several metallic objects, such as desks, chairs, computer towers and monitors, we deploy $32$ sensors to cover a $67 m^2$ area ($27$ sensors along the perimeter, $5$ in internal positions). The nodes along the perimeter are placed on podiums at a height of one meter from the floor, while the other five are positioned on desks. They transmit on $5$ different channels, \emph{i.e.} $c \in \{11,15,18,22,26\}$. In this environment, we conduct several tests with two, three, and four people simultaneously walking at constant speed along a pre-defined path. At first, people enter the area one after the other and walk along the path in the same direction, exiting the area again one after the other. Later, people enter the area again one after the other, but this time they walk in opposite direction, so that their trajectories intersect multiple times.

\section{Results} \label{sec:tests_results}
\subsection{Evaluation metrics}
\label{sec:evaluation_metrics}

\subsubsection{Cardinality error}
The cardinality error $\epsilon_c$ is calculated as the fraction of frames during the test in which the number of people moving in the monitored area and the estimated number of confirmed tracks differ.

\subsubsection{Tracking accuracy}
To evaluate the tracking accuracy, we use the optimal mass transfer (OMAT) metric, introduced in \cite{original_OMAT} and defined as:
\begin{equation}\label{E:OMAT_metric}
    {\epsilon_{O}(\mathcal{T},\mathcal{Z})} = \frac{1}{\left| \mathcal{T} \right|} \min_{\upsilon \in \Upsilon} \left( {\sum_{i=1}^{|{\mathcal{T}}|}} {\sum_{j=1}^{|{\mathcal{Z}}|}} {\upsilon}_{i,j} \| {\mathbf{t}_{i}-\mathbf{z}_{j}} \|^q \right)^{\frac{1}{q}},
\end{equation}
in which $\Upsilon$ is the set of all the possible permutations $\upsilon$ between the set of estimated targets $\mathcal{T}$ and the set of real targets $\mathcal{Z}$, and $q$ is the order of the metric (we set $q=2$). Thus, the OMAT error $\epsilon_O$ is equal to the root mean square error (RMSE) of the best possible association between estimated and real targets.

\subsubsection{OSPA metric}
Differently than the OMAT metric defined in (\ref{E:OMAT_metric}), the optimal subpattern assignment (OSPA) metric \cite{OSPA_metric} includes an additional term, \emph{i.e.}, a constant $g$ measured in meters, which penalizes the cardinality error. When $\mathcal{T}$ is the set of estimated targets, $\mathcal{Z}$ the set of real targets, and $\left| \mathcal{T} \right| \leq \left| \mathcal{Z} \right|$, the OSPA metric $\epsilon_{P}^{(g)}(\mathcal{T},\mathcal{Z})$ can be calculated as:
\begin{equation}\label{E:OSPA_metric}
\begin{split}
    {\epsilon_{P}^{(g)}(\mathcal{T},\mathcal{Z})} = \\
\left( \frac{1}{|\mathcal{Z}|} \min_{\upsilon \in \Upsilon} {\sum_{i=1}^{|{\mathcal{T}}|}} d^{(g)} \left( t_{i},z_{\upsilon \left( i \right)} \right)^{q} + g^{q} \left( \left| \mathcal{Z} \right| - \left| \mathcal{T} \right| \right) \right)^{\frac{1}{q}},
\end{split}
\end{equation}
where $d^{(g)} \left( t,z \right) = \min \{ d \left( t,z \right),g \}$ and $d$ is the Euclidean distance between the estimated and real position of a target. We set $q=2$. When $\left| \mathcal{T} \right| > \left| \mathcal{Z} \right|$, the OSPA metric is calculated as $\epsilon_{P}^{(g)}(\mathcal{Z},\mathcal{T})$, \emph{i.e.}, as in (\ref{E:OSPA_metric}) but inverting $\mathcal{T}$ and $\mathcal{Z}$ in it.

\subsubsection{Tracking consistency}
Since in some of the tests the trajectories of the targets intersect, we define a metric to evaluate the capability of the DFL system to avoid losing track of targets during and after their intersection. To this purpose, we use the 95\%-percentile $Q_{95}$ of the OMAT errors of the estimated tracks. A low $Q_{95}$ indicates that even with intersecting trajectories and noisy RTI images the system does not lose track of the targets.

\subsection{Experimental results}
\label{sec:experimental_results}

\subsubsection{Processing time}
The maximum, $\max{[T_p]}$, and average, $E[T_p]$, processing time of the multiple target tracking algorithms are measured on a laptop having a $2.50$ GHz Intel\textsuperscript{\textregistered} Core\textsuperscript{TM} i5-2450P processor and $8.0$ GB of RAM memory. Table \ref{T:proc_time_results} lists $\max{[T_p]}$ and $E[T_p]$ for the two considered nearest neighbor approaches, GNN and SNN. For comparison, the table includes also $\max{[T_p]}$ and $E[T_p]$ for the multiple hypothesis tracking (MHT) method \cite{ReidMHT,Blackman_MHT}. Both the methods are well below the limit set by the length of an entire TDMA communication cycle on one channel (see Section \ref{sec:hardware}), which is approximately $100$ ms for a network composed of about $30$ nodes. The results show that the SNN method is faster than the GNN method. On the other hand, the MHT method does not fulfill the real-time requirement, especially when the trajectories of the targets intersect. In this situation, the MHT method starts creating a tree of hypothesis for the possible paths followed by the targets based on the available observations, postponing the decision till following RTI images which are more easy to be interpreted. Moreover, the processing time of the MHT method increases considerably with the number of targets to be tracked due to the increased size of the hypothesis tree. Both $\max{[T_p]}$ and $E[T_p]$ increase with the number of targets to be tracked also for the GNN and SNN methods, due to the higher number of voxels going through the clustering process and the higher number of observations that have to be associated to the existing tracks. However, both methods cope much better than MHT with the increased complexity of the tracking problem.

\begin{figure}[t!]
  \begin{center}
    \mbox{
      \subfigure[\quad 2 targets - open environment]{\includegraphics[width=\columnwidth]{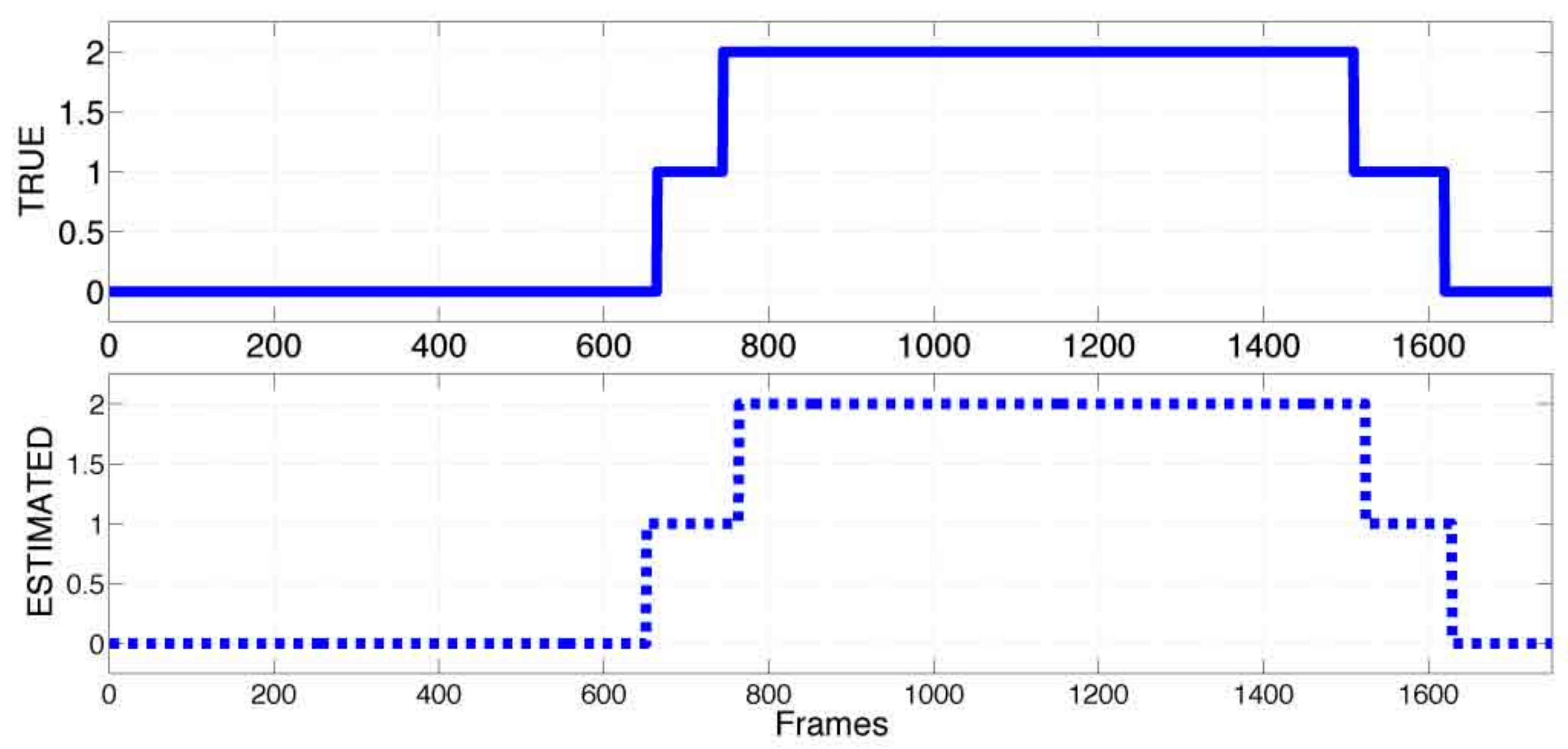}} \quad
      }
    \mbox{
      \subfigure[\quad 3 targets - office environment]{\includegraphics[width=\columnwidth]{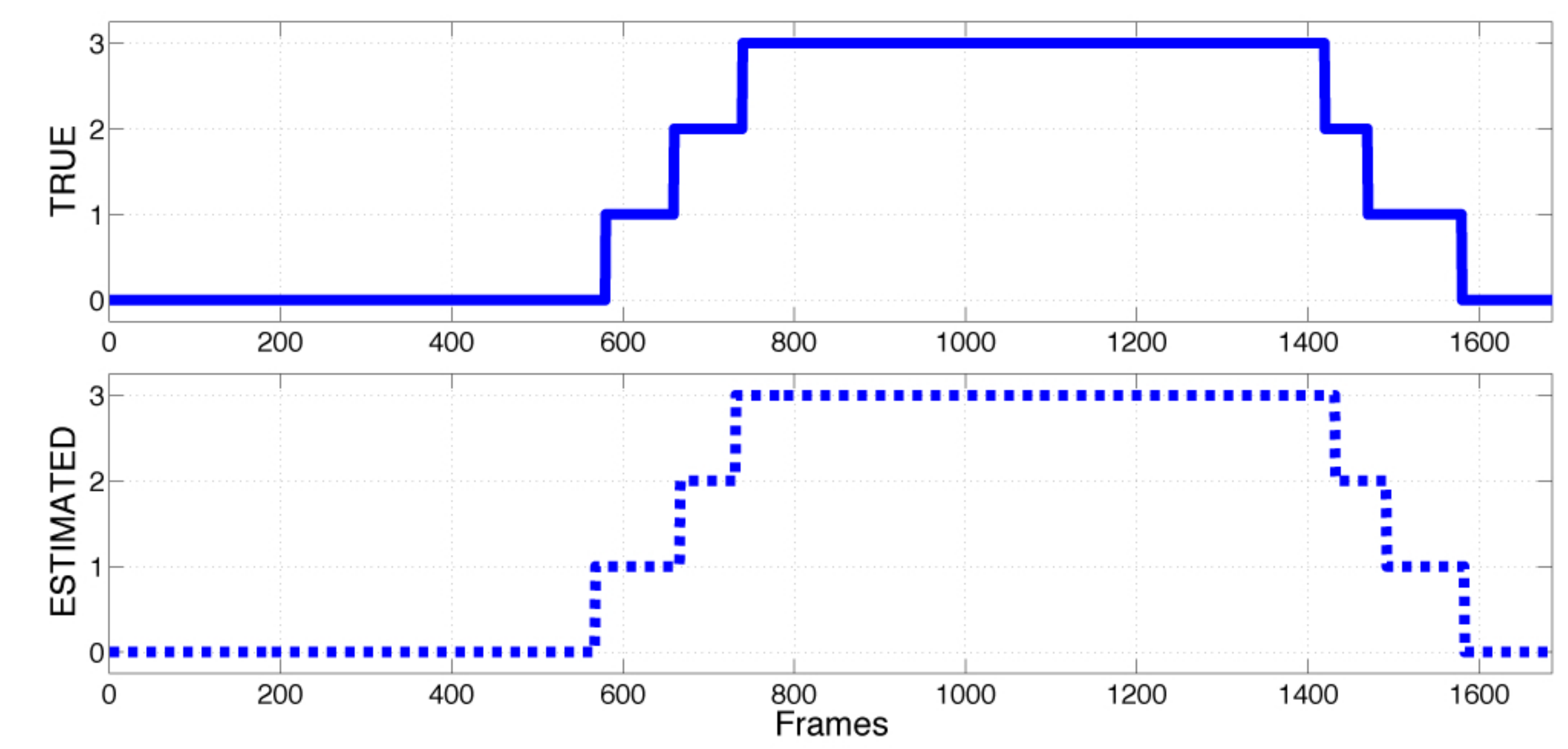}} \quad
      }
    \mbox{
      \subfigure[\quad 4 targets - office environment]{\includegraphics[width=\columnwidth]{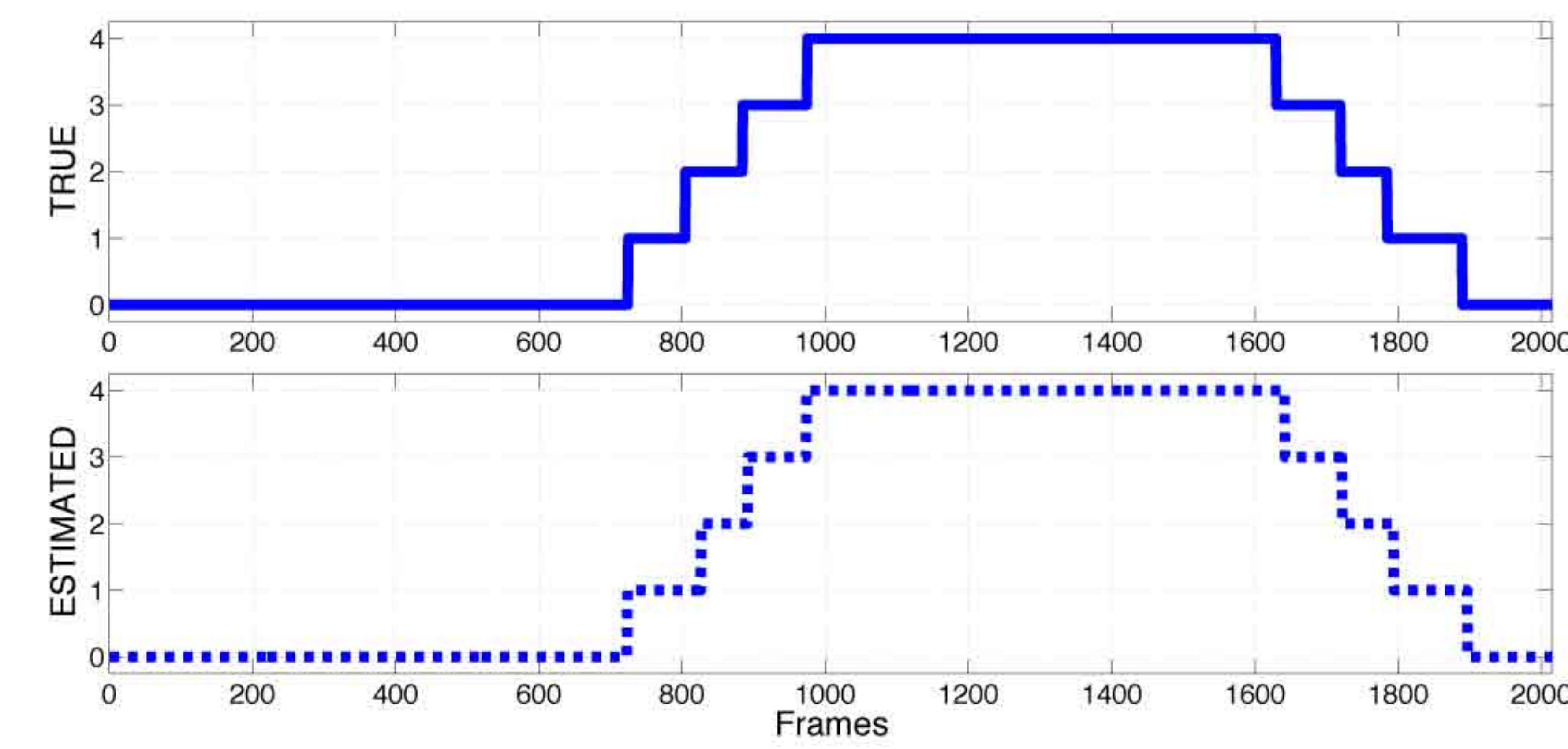}} \quad
      }   
    \caption{In (a)-(c), the comparison between the true numer of targets in the monitored area and the number of targets estimated by the DFL system during the tests carried out in the office environment with targets having separated trajectories.}
    \label{F:cardinality_figure}
  \end{center}
\end{figure}

\subsubsection{Cardinality error}
The cardinality error ($\epsilon_c$) and the accuracy ($\epsilon_O$) and consistency ($Q_{95}$) of the tracking obtained by applying the methods presented in this paper are listed in Table \ref{T:fade_level_tracking_results}. The cardinality error measured in all the experiments is limited only to a few frames before the actual entrance of the targets in the monitored area and a few frames after their actual exit, as shown in Figure \ref{F:cardinality_figure}. This depends on the confirmation and deletion rules discussed in Section \ref{sec:tracksconfdel}.

During the experiments we observed that the delay in detecting the entrance of a target in the monitored area due to the confirmation rule can sometimes be compensated by the use of multiple frequency channels. Deep-fade links show variation of the RSS even when the person is at some position far away from the link line. When a new target approaches the entrance region, the RTI images start showing a new blob. If this blob receives multiple successive confirmations, the presence of a new target is detected with a small anticipation with respect to its actual entrance in the monitored area. This effect can be observed in Figures \ref{F:cardinality_figure}(a) and \ref{F:cardinality_figure}(b) at the entrance of the first target.

\begin{figure}[t!]
  \begin{center}
    \includegraphics[width=\columnwidth,height=5.50cm]{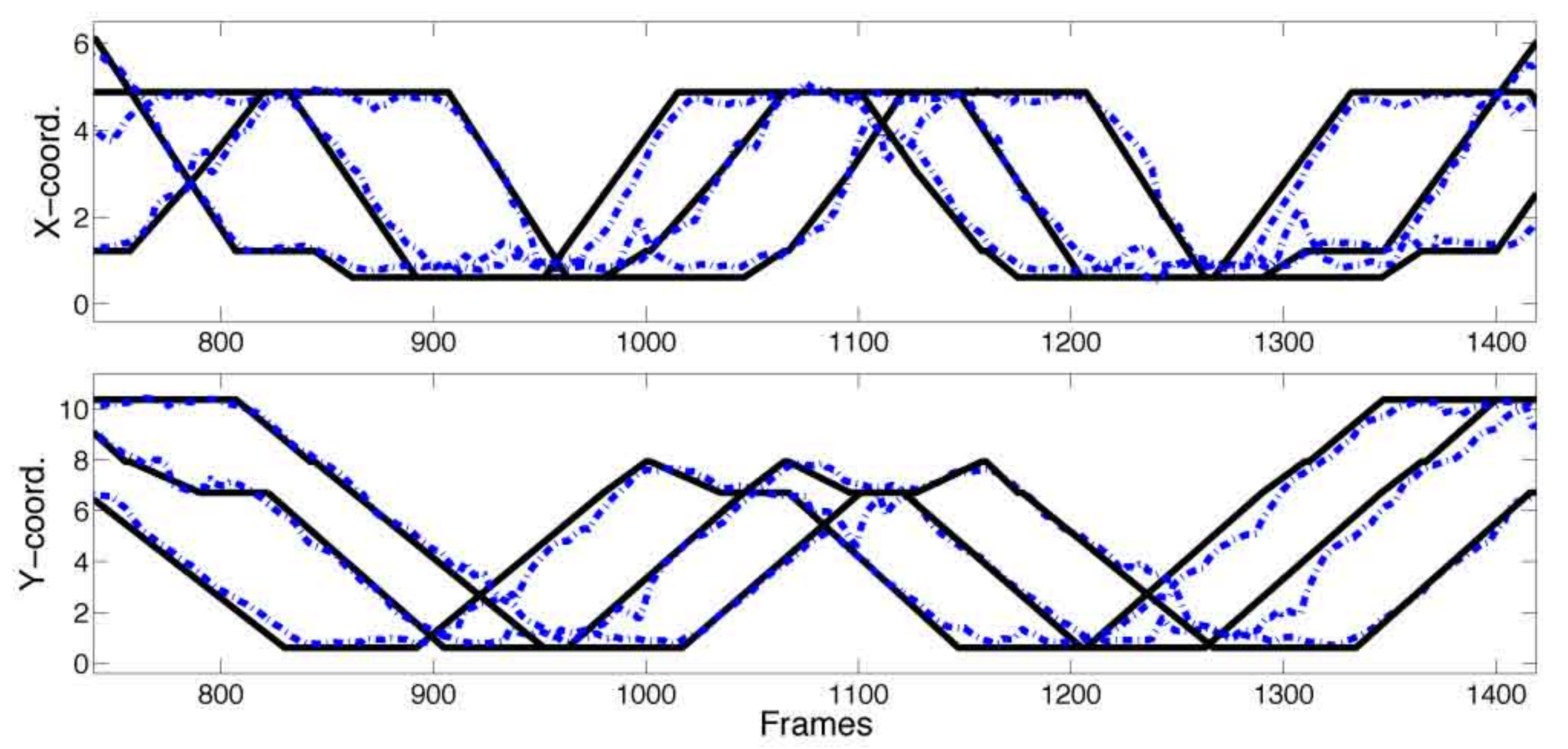}
    \caption{The real (solid lines) and estimated (dashed lines) coordinates of the targets during a test carried out in the office environment. Despite multiple intersections, $\epsilon_O = 0.44$ m and $Q_{95} = 0.99$ m.}
    \label{F:trajectories_figure}
  \end{center}
\end{figure}

\subsubsection{Tracking accuracy and consistency}
For what concerns the tracking accuracy, the largest average OMAT error, $\epsilon_O = 0.55$ m, is measured when four targets with intersecting trajectories are tracked in the office environment, where multipath propagation of the radio signals is predominant. On the other hand, the largest OMAT errors measured in the open environment are $\epsilon_O = 0.33$ m and $\epsilon_O = 0.30$ m, respectively.

In the apartment environment, the difference in the $\epsilon_O$ and $Q_{95}$ values measured in the two tests carried out is due to the paths covered by the two people, which were more separated in the second experiment than in the first one. In fact, the spots where the two tracks intersect have a huge impact on the consistency of the tracking accuracy. Due to the particular positioning of the sensors in an environment with walls, various objects, and furniture, certain areas are covered by a higher number of anti-fade links (which favor the localization effort), whereas in some others the number of deep-fade links (less reliable for localization) is larger \cite{kaltiokallio2012b}. For this reason, the amount of noise found in the RTI images formed when the two trajectories intersect is different in each test, and this explains the slower convergence of the estimated tracks to the real ones after the intersections. Despite this, the average OMAT error measured during the two tests is approximately $0.30$ m, both with the GNN and SNN methods.

\begin{center}
\begin{table*}[t!]
    \caption{$\epsilon_c$, $\epsilon_O$ and $Q_{95}$ for the GNN and SNN methods} %title of the table
        \centering
        \footnotesize
        \begin{tabular}{c c c c | c c | c c }
        \hline\hline %insert double horizontal line
        &  &  &  & \multicolumn{2}{c|}{GNN method} & \multicolumn{2}{c}{SNN method} \\
        \hline
        Targets & Environment & Intersections (\#) & $\epsilon_c$ & $\epsilon_O$ [m] & $Q_{95}$ [m] & $\epsilon_O$ [m] & $Q_{95}$ [m]\\
        \hline
        2 & open        & NO             & 0.001         & 0.32   & 0.71          & 0.33   & 0.72   \\ %body of the table
        2 & open        & YES (9)        & 0.001         & 0.32   & 0.74          & 0.33   & 0.77   \\ %body of the table
        \hline %insert single horizontal line
        2 & apartment   & YES (4)        & 0.001         & 0.30   & 0.71          & 0.30   & 0.75   \\ %body of the table
        2 & apartment   & YES (5)        & 0.002         & 0.26   & 0.68          & 0.27   & 0.71   \\ %body of the table
        \hline %insert single horizontal line
        2 & office      & NO             & 0.029         & 0.38   & 0.75          & 0.38   & 0.76   \\ %body of the table
        2 & office      & YES (5)        & 0.023         & 0.44   & 0.96          & 0.45   & 1.01   \\ %body of the table
        3 & office      & NO             & 0.016         & 0.37   & 0.66          & 0.37   & 0.67   \\ %body of the table
        3 & office      & YES (6)        & 0.032         & 0.44   & 1.05          & 0.46   & 1.07   \\ %body of the table
        4 & office      & NO             & 0.027         & 0.44   & 0.94          & 0.45   & 0.96   \\ %body of the table
        4 & office      & YES (9)        & 0.029         & 0.54   & 1.16          & 0.55   & 1.26   \\ %body of the table
        \hline %inserts single line
        \end{tabular}
        \label{T:fade_level_tracking_results}
\end{table*}
\end{center}

\begin{figure}[t!]
  \begin{center}
    \includegraphics[width=\columnwidth,height=4.75cm]{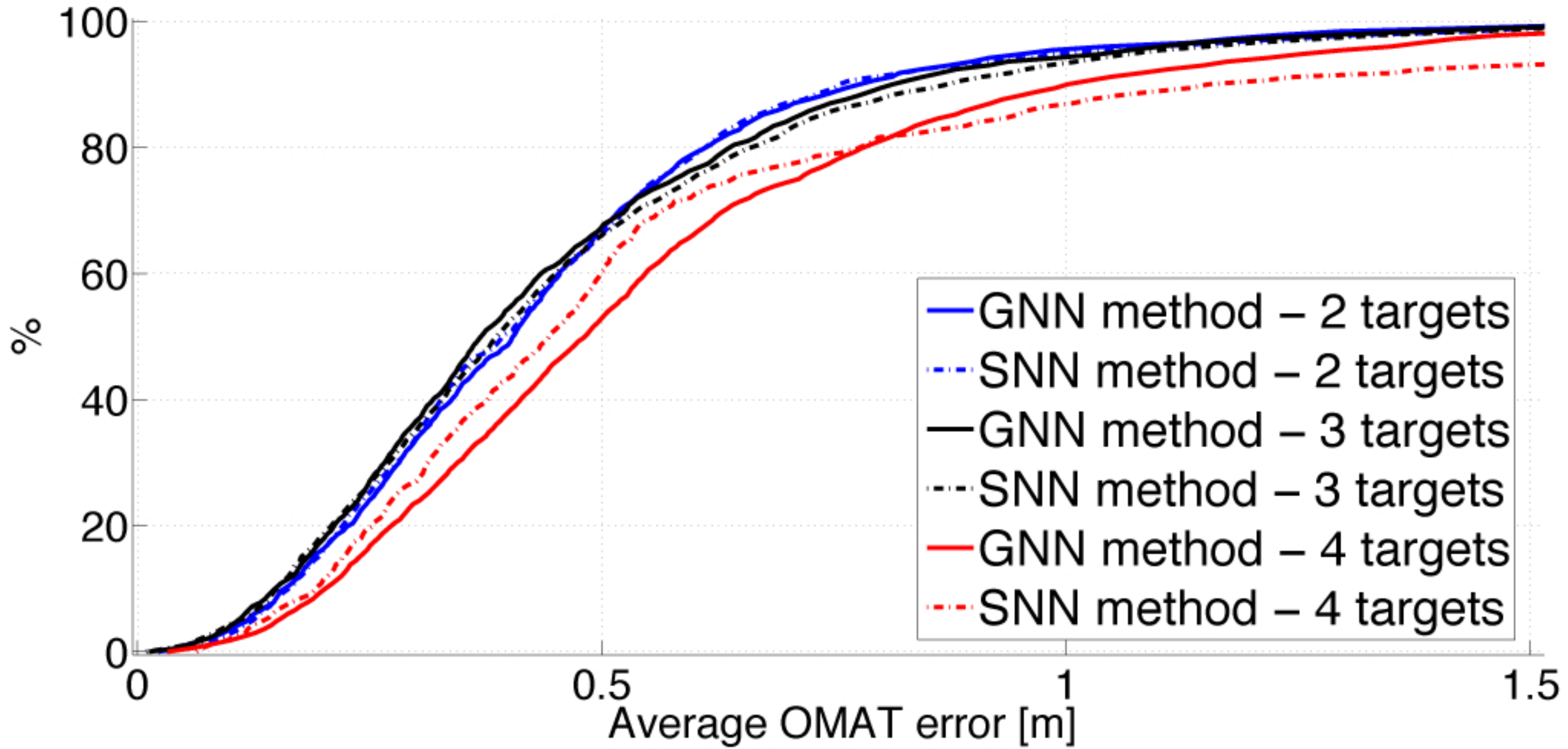}
    \caption{The cumulative distribution functions of the average OMAT error obtained with the GNN (solid lines) and SNN (dashed lines) methods in the tests carried out in the office environment with targets having intersecting trajectories.}
    \label{F:CDFs_figure}
  \end{center}
\end{figure}

In the office environment, the average OMAT error measured with two targets is slightly higher than in the other two environments, in both the situations of separated ($\epsilon_O = 0.38$ m) and intersecting ($\epsilon_O = 0.45$ m) trajectories. The difference is due to the fact that this environment is the most challenging for DFL due to the presence of multiple desks, chairs, computer towers and monitors. Nevertheless, the largest $\epsilon_O$ and $Q_{95}$ measured with 4 targets are $0.45$ m and $0.96$ m with separated trajectories and $0.55$ m and $1.26$ m with intersecting trajectories. These results show that our DFL system is capable of accurately and consistently tracking up to $4$ targets in a challenging indoor environment where multipath propagation is predominant. Figure \ref{F:trajectories_figure} shows the real and estimated trajectories of the test conducted in the office environment with three targets having intersecting trajectories. The cumulative distribution functions (CDFs) of $\epsilon_O$ obtained with the GNN and SNN methods in the tests carried out in the office environment with intersecting trajectories are shown in Figure \ref{F:CDFs_figure}. Since the GNN method finds the optimal solution of the DAP at each frame, the improvement in $Q_{95}$ measured with this method becomes more consistent with more noisy images, \emph{i.e.} when four intersecting targets are tracked.

\begin{table}[t!]
    \caption{Average OSPA errors for the tests in the office environment} %title of the table
        \centering
        \footnotesize
        \begin{tabular}{c c | c c c}
        \hline\hline %insert double horizontal line
        &  &  \multicolumn{3}{c}{Average OSPA error [m]} \\
        \hline
        Targets & Intersections & $g=1$ & $g=2.5$ & $g=5$ \\
        \hline
        2  & no   & 0.41 & 0.45 & 0.51   \\ %body of the table
        2  & yes  & 0.46 & 0.51 & 0.56   \\ %body of the table
        3  & no   & 0.39 & 0.41 & 0.43   \\ %body of the table
        3  & yes  & 0.48 & 0.59 & 0.75   \\ %body of the table
        4  & no   & 0.47 & 0.52 & 0.59   \\ %body of the table
        4  & yes  & 0.57 & 0.69 & 0.83   \\ %body of the table
        \hline %inserts single line
        \end{tabular}
        \label{T:OSPA_table}
\end{table}

Table \ref{T:OSPA_table} lists the average OSPA errors for the tests carried out in the office environment for different values of the cardinality penalty $g$. Despite the fact that no assumption is made on the number of targets found in the monitored area, due to the very small fraction of frames in which $|{\mathcal{T}}| \ne |\mathcal{Z}|$, the average OSPA errors increase by a very small amount even when the cardinality penalty is set to a very high value ($g=5$).

\begin{figure}[t]
  \begin{center}
    \includegraphics[width=\columnwidth,height=5.25cm]{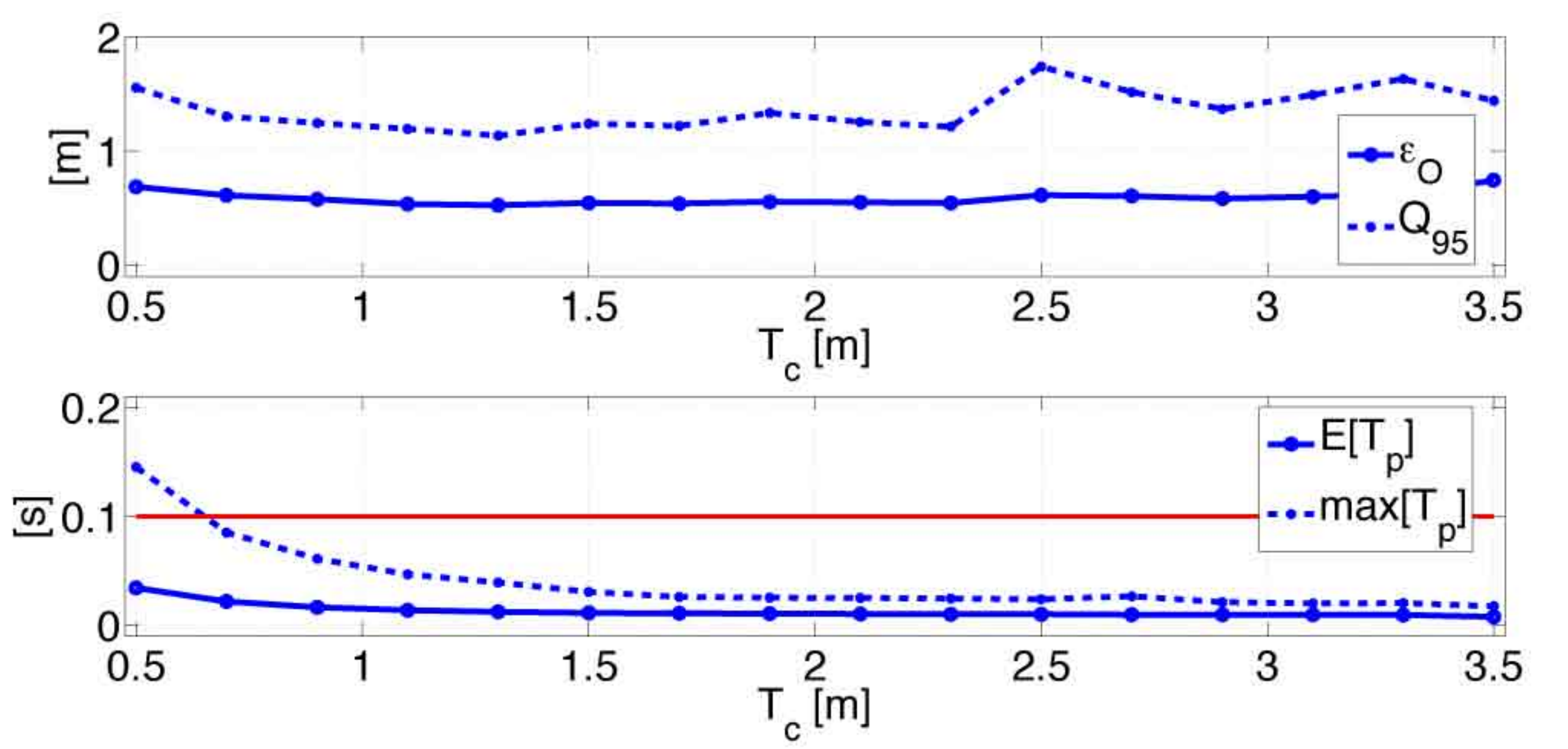}
    \caption{$\epsilon_O$, $Q_{95}$, $\max{[T_p]}$, and $E[T_p]$ measured with the GNN method for different values of $T_c$ for the test carried out in the office environment with $4$ targets having intersecting trajectories. The solid line in the image at the bottom represents the real-time limit for the processing time.}
    \label{F:sensitivity_figure}
  \end{center}
\end{figure}

\subsubsection{Sensitivity analysis}
Figure \ref{F:sensitivity_figure} shows $\epsilon_O$, $Q_{95}$, $\max{[T_p]}$, and $E[T_p]$ measured with the GNN method for different values of the clustering threshold $T_c$ (see Section \ref{sec:clustering}) for the test carried out in the office environment with $4$ targets having intersecting trajectories. $T_c$ determines the average size of the formed clusters, and ultimately their number. The results show that the tracking accuracy of the system is consistent for different values of $T_c$, with $\epsilon_O$ taking values from $0.54$ m to $0.74$ m. Related to the processing time, when $T_c$ takes small values the system is not capable of processing the RTI images in real-time. This is due to the large number of formed clusters, which increases the time required for solving the data association problem (see Section \ref{sec:targets_tracking}). Both $\max{[T_p]}$ and $E[T_p]$ decrease when $T_c$ takes larger values. Similar results were obtained while considering all the other tests.

\section{Related Work} \label{sec:related_work}
\begin{center}
\begin{table*}[t!]
    \caption{Comparison table} %title of the table
        \centering
        \footnotesize
        \begin{tabular}{>{\centering\arraybackslash}m{2.0cm} >{\centering\arraybackslash}m{1.0cm} >{\centering\arraybackslash}m{1.5cm} >{\centering\arraybackslash}m{2.5cm} >{\centering\arraybackslash}m{2cm} >{\centering\arraybackslash}m{1.5cm} >{\centering\arraybackslash}m{4.5cm}}
        \hline\hline %insert double horizontal line
        \\ [-2.0ex]
        Work & Targets & Intersections & Environment & Accuracy [m] & Latency [s] & Assumptions \\
        \\ [-2.0ex]
        \hline
        \\ [-2.0ex]
        \shortstack[m]{Zhang \emph{et al.} \\ 2009} & 2 & no  & Uncluttered indoor & 1.08 & 2.00 & \shortstack[m]{Targets at least 5 m apart.} \\
        \\ [-2.0ex]
        \hline %insert single horizontal line
        \\ [-2.0ex]
        \shortstack[m]{Zhang \emph{et al.} \\ 2011} & 2 & no  & Uncluttered indoor & 0.98 & 0.26 & \shortstack[m]{Targets at least 2 m apart.} \\
        \\ [-2.0ex]
        \hline %insert single horizontal line
        \\ [-2.0ex]
        \shortstack[m]{Wilson and Patwari \\ 2011}  & 2 & no  & \shortstack[m]{Bookstore \\ Through wall} & \shortstack[m]{0.84 \\ 1.10} & \shortstack[m]{Not in real-time} & \shortstack[m]{$|\mathcal{T}|$ fixed and known a priori} \\
        \\ [-2.0ex]
        \hline %insert single horizontal line
        \\ [-2.0ex]
        \shortstack[m]{Thouin \emph{et al.} \\ 2011} & 4 & no  & Uncluttered outdoor & \shortstack[m]{0.63 \\ 0.77 \\ 0.96} & \shortstack[m]{7.6 \\ 1.7 \\ 0.048 (Real-time)} & \shortstack[m]{$|\mathcal{T}|$ fixed and known a priori.} \\
        \\ [-2.0ex]
        \hline %insert single horizontal line
        \\ [-2.0ex]
        \shortstack[m]{Nannuru \emph{et al.} \\ 2012} & 3 & no  & \shortstack[m]{Uncluttered indoor \\ Office} & 0.80 & \shortstack[m]{Real-time if \\ $N_p \le 100$} & \shortstack[m]{$|\mathcal{T}|$ fixed and known a priori. \\ Frequent $\epsilon_c$ with varying $N_t$.} \\
        \\ [-2.0ex]
        \hline %insert single horizontal line
        \\ [-2.0ex]
        \shortstack[m]{Bocca \emph{et al.} \\ 2012} & 4 & yes  & \shortstack[m]{Uncluttered indoor \\ Apartment \\ Office} & 0.55 & \shortstack[m]{Real-time} & \\
        \\ [-2.0ex]
        \hline %insert single horizontal line
        \end{tabular}
        \label{T:comparison_table}
\end{table*}
\end{center}

In this section, we review the previous works related to multiple target tracking with RF sensor networks. Table \ref{T:comparison_table} summarizes the characteristics of these systems. In \cite{Zhang_2009}, the authors present a clustering algorithm to form clusters of those links whose RSS is affected by the same object. In this work, the nodes are positioned on the ceiling of an open indoor environment. The RSS measurements of each identified cluster are then processed separately to localize the targets. The results show that this approach achieves a $1.08$ m RMSE with two targets. However, to achieve these results, the probabilistic cover algorithm introduced in the paper requires that the two targets are separated by at least 5 m. Moreover, the latency of the system is approximately 2 s. In \cite{Zhang_2011}, the monitored area is divided into different sections, each of which is covered by three nodes communicating on a different frequency channel and positioned on the ceiling as to form a triangle. A support vector regression model is applied to locate the targets. The results show that the system can detect and locate two targets when these are positioned in different triangles, \emph{i.e.} at least $2$ m apart, with a $0.98$ m RMSE. With this system, the latency is reduced to 0.26 s.

The work in \cite{wilson11fade} presents the results of tests carried out in a cluttered book store and in a through-wall scenario with $2$ targets having separated trajectories. The authors use a particle filter \cite{Ristic_PF} to track the two targets, measuring an RMSE of $0.84$ m in the book store and of $1.1$ m in the through-wall scenario. The number of targets is assumed to be known a priori. Due to the computational complexity of the particle filter method, localization and tracking are not performed in real-time.

In \cite{Thouin_2011}, the authors introduce a novel measurement model that assumes that the attenuation in RSS due to the simultaneous presence of multiple targets on the link line is approximately equal to the sum of the attenuations due by the single targets. This additive model is then applied in \cite{MTT_tracking_Nannuru2011} and \cite{MTT_tracking_Nannuru2012}. The tests reported in \cite{MTT_tracking_Nannuru2011} are conducted in an outdoor uncluttered environment with up to four targets having separated trajectories. During each test, the number of targets is assumed to be fixed and known a priori. For tracking, three types of PF algorithms are used. The initial set of particles for each target is drawn from a Gaussian distribution centered around the real target location. The results show the existing trade-off between the tracking accuracy (which depends on the number of particles used per target) and the processing time (which depends on the computational complexity of the tracking algorithms). With four targets, the most accurate tracking algorithm achieves $0.63$ m average error with $500$ particles per target, requiring in average $7.6$ s per time step. When $50$ particles per target are used, the most accurate tracking algorithm achieves $0.77$ m error, requiring $1.7$ s per time step. On the other hand, the fastest algorithm of the three taken into consideration achieves $0.96$ m error in $48$ ms per time step. In comparison, our system achieves lower average error ($0.55$ m) requiring on average $13.3$ ms per time step.

In \cite{MTT_tracking_Nannuru2012}, the tests are carried out in an open indoor environment and in an office environment with up to three targets having separated trajectories. The authors use different particle filters to track the targets. The filters perform better in average when the particles are initialized according to a Gaussian distribution centered at the true targets locations, which in this case are assumed to be known a priori. The performance of the filters decays when the particles are initialized according to a uniform distribution within the observation region. When the number of targets is assumed to be fixed and known a priori, the most accurate tracking algorithm achieves a RMSE of $0.30$ m, $0.72$ m, and $0.80$ m with one, two, and three targets, respectively. However, when the number of targets is varying, the applied algorithms are prone to mis-estimate the number of targets in several time steps. In the tests with two targets, this increases the average OSPA error to $1.35$ m when the cardinality penalty $g = 5$. In comparison, our system achieves an average error of $0.45$ m, $0.46$ m, and $0.55$ m with two, three, and four targets, and a maximum OSPA error of $0.83$ with four targets when $g = 5$.

\section{Conclusion} \label{sec:paper_conclusion}
This paper presents a RF sensor network capable of tracking in real-time multiple targets simultaneously moving in an area where low-power wireless transceivers are deployed. The system does not require people to be tracked to participate in the localization task by carrying any radio device or RFID tag. Instead, the RSS measurements collected on the links forming the mesh network on multiple frequency channels are processed to form RTI images, \emph{i.e.}, images of the change in the propagation field due to the presence of people in the area. Using machine vision methods adapted to RTI, we process the RTI images in real-time to detect and track the blobs corresponding to real targets. In this work, we address the situation in which the targets have intersecting trajectories. Moreover, we apply computationally light-weight methods that can be executed in real-time. Furthermore, during the tests the number of targets is varying and is not assumed to be known a priori.

We conduct experiments in three different indoor environments, \emph{i.e.} an open environment with no obstructions nor objects, a one-bedroom apartment with internal walls, furniture and various objects, and a heavily cluttered office environment. In the tests, up to four targets with intersecting trajectories are tracked. The results show that our DFL system is able to correctly estimate the number of targets found in the monitored area and to accurately track them in real-time in all three environments, also when their trajectories intersect. The measured tracking RMSE ranges from $0.27$ m (achieved in the apartment environment with two targets having intersecting trajectories) to $0.52$ m (achieved in the office environment with four targets having intersecting trajectories).

In future work, we will address other challenging tracking situations, such as people entering the monitored area side-by-side and then splitting and moving in different directions. In addition, whenever two (or more) targets converge and then separate again, the system has a high probability of not keeping the correct track-to-blob association. To overcome this limitation, we will equip the targets with an active badge and use the AGAPE method presented in \cite{Zhao_directionality} to maintain the correct track-to-blob association at all times.

\section*{Acknowledgments}
This work is supported in part by the US National Science Foundation Grants \#0748206 and \#1035565 and by the Finnish Funding Agency for Technology and Innovation (TEKES). The authors thank Brad Mager for his support in setting up the experiments.
\addcontentsline{toc}{section}{Acknowledgment}

%==================================================================================================%
%  Bibliography
%==================================================================================================%
%\bibliographystyle{IEEEtran}
%\bibliography{overall}

\begin{thebibliography}{44}
%-----------------------------------------------------%
%-------------------- INTRODUCTION -------------------%
%1
\bibitem{wilson09a} J. Wilson and N. Patwari, “Radio tomographic imaging with wireless networks,” IEEE Trans. Mobile Computing, vol. 9, no. 5, pp. 621–632, May 2010, appeared online 8 January 2010.
%2
\bibitem{kanso09b} M. A. Kanso and M. G. Rabbat, “Compressed RF tomography for wireless sensor networks: Centralized and decentralized approaches,” in 5th IEEE Intl. Conf. on Distributed Computing in Sensor Systems (DCOSS-09), Marina Del Rey, CA, June 2009.
%3
\bibitem{chen11sequential} X. Chen, A. Edelstein, Y. Li, M. Coates, M. Rabbat, and M. Aidong, “Sequential monte carlo for simultaneous passive device-free tracking and sensor localization using received signal strength measurements,” in ACM/IEEE Information Processing in Sensor Networks (IPSN), April 2011.
%4
\bibitem{kaltiokallio2011} O. Kaltiokallio and M. Bocca, “Real-Time Intrusion Detection and Tracking in Indoor Environment through Distributed RSSI Processing,” in 2011 IEEE 17th Intl. Conf. Embedded and Real-Time Computing Systems and Applications (RTCSA), 2011, pp. 61–70.
%5
\bibitem{patwari10c} N. Patwari and J. Wilson, “RF sensor networks for device-free localization and tracking,” Proceedings of the IEEE, vol. 98, no. 11, pp. 1961–1973, Nov. 2010.
%6
\bibitem{kaltiokallio2012b} O. Kaltiokallio, M. Bocca, and N. Patwari, “Follow @grandma: Long-term device-free localization for residential monitoring,” in 7th IEEE International Workshop on Practical Issues in Building Sensor Network Applications, October 2012.
%7
\bibitem{breathing_Patwari} N. Patwari, J. Wilson, S. Ananthanarayanan, S. K. Kasera, and D. R. Westenskow, “Monitoring breathing via signal strength in wireless networks,” ArXiv.org, vol. abs/1109.3898, 2011.
%8
\bibitem{whitehouse_2012} T. Hnat, E. Griffiths, R. Dawson, and K. Whitehouse, “Doorjamb: unobtrusive room-level tracking of people in homes using doorway sensors,” in Proceedings of the 10th ACM Conference on Embedded Networked Sensor Systems (SenSys 2012), November 2012.
%9
\bibitem{hashemi94} H. Hashemi, “A study of temporal and spatial variations of the indoor radio propagation channel,” in PIMRC-94, vol. 1, Sep 1994, pp. 127–134.
%10
\bibitem{ghaddar2004human} M. Ghaddar, L. Talbi, and T. Denidni, “Human body modelling for prediction of effect of people on indoor propagation channel,” Electronics letters, vol. 40, p. 25, 2004.
%11
\bibitem{wilson10see} J. Wilson and N. Patwari, “See through walls: motion tracking using variance-based radio tomography networks,” IEEE Trans. Mobile Computing, vol. 10, no. 5, pp. 612–621, May 2011, appeared online 23 September 2010.
%12
\bibitem{kaltiokallio2012a} O. Kaltiokallio, M. Bocca, and N. Patwari, “Enhancing the accuracy of radio tomographic imaging using channel diversity,” in 9th IEEE International Conference on Mobile Ad hoc and Sensor Systems, October 2012.
%13
\bibitem{wilson11fade} J. Wilson and N. Patwari, “A fade level skew-Laplace signal strength model for device-free localization with wireless networks,” IEEE Trans. Mobile Computing, appeared online 12 May 2011.
%14
\bibitem{MTT_videos} M. Bocca. [Online].\\Available: https://sites.google.com/site/boccamaurizio/research
%15
\bibitem{Zhang_2009} D. Zhang and L. M. Ni, “Dynamic clustering for tracking multiple transceiver-free objects,” in Proceedings of the 2009 IEEE International Conference on Pervasive Computing and Communications, ser. PERCOM ’09. Washington, DC, USA: IEEE Computer Society, 2009, pp. 1–8.
%16
\bibitem{Zhang_2011} D. Zhang, Y. Liu, and L. Ni, “Rass: A real-time, accurate and scalable system for tracking transceiver-free objects,” in 2011 IEEE International Conference on Pervasive Computing and Communications (PerCom), March 2011, pp. 197 –204.
%17
\bibitem{MTT_tracking_Nannuru2012} S. Nannuru, Y. Li, Y. Zeng, M. Coates, and B. Yang, “Radio frequency tomography for passive indoor multi-target tracking,” IEEE Transactions on Mobile Computing, vol. PP, no. 99, p. 1, 2012.
%18
\bibitem{MTT_tracking_Nannuru2011} S. Nannuru, Y. Li, M. Coates, and B. Yang, “Multi-target device-free tracking using radio frequency tomography,” in 2011 7th International Conference on Intelligent Sensors, Sensor Networks and Information Processing (ISSNIP), dec. 2011, pp. 508 –513.
%19
\bibitem{Thouin_2011} F. Thouin, S. Nannuru, and M. Coates, “Multi-target tracking for measurement models with additive contributions,” in 2011 Proceedings of the 14th International Conference on Information Fusion (FUSION), July 2011, pp. 1 –8.
%20
\bibitem{tidongleantenna} Texas Instruments. Small Size 2.4 GHz PCB antenna. [Online]. Available: www.ti.com/lit/an/swra117d/swra117d.pdf
%21
\bibitem{patwari08b} N. Patwari and P. Agrawal, “Effects of correlated shadowing: Connectivity, localization, and RF tomography,” in IEEE/ACM Int’l Conf. on Information Processing in Sensor Networks (IPSN’08), April 2008, pp. 82–93.
%22
\bibitem{agrawal09} P. Agrawal and N. Patwari, “Correlated link shadow fading in multi-hop wireless networks,” IEEE Trans. Wireless Commun., vol. 8, no. 8, pp. 4024–4036, Aug. 2009.
%23
\bibitem{zhao11noise} Y. Zhao and N. Patwari, “Noise reduction for variance-based device-free localization and tracking,” in 8th IEEE Conference on Sensor, Mesh and Ad Hoc Communications and Networks (SECON’11), June 2011.
%24
\bibitem{davies2012computer} E. Davies, Computer and Machine Vision: Theory, Algorithms, Practicalities, ser. Academic Press. Elsevier Science and Technology, 2012. [Online]. Available: http://books.google.com/books?id=AhVjXf2yKtkC
%25
\bibitem{kmeans} J. B. MacQueen, “Some methods for classification and analysis of multivariate observations,” in Proc. of the 5th Berkeley Symposium on Mathematical Statistics and Probability, L. M. L. Cam and J. Neyman, Eds., vol. 1. University of California Press, 1967, pp. 281–297.
%26
\bibitem{hastie2009elements} T. Hastie, R. Tibshirani, and J. Friedman, The Elements of Statistical Learning: Data Mining, Inference, and Prediction, Second Edition, ser. Springer Series in Statistics. Springer, 2009.
%27
\bibitem{blackman99} S. Blackman and R. Popoli, Design and Analysis of Modern Tracking Systems. Artech House Publishers, 1999.
%28
\bibitem{Ristic_PF} B. Ristic, S. Arulampalam, and N. Gordon, Beyond the Kalman Filter: Particle Filters for Tracking Applications. Artech House, Jan. 2004.
%29
\bibitem{JPDA_Bar_Shalom} Y. Bar-Shalom, F. Daum, and J. Huang, “The probabilistic data association filter,” IEEE Control Systems Magazine, vol. 29, no. 6, pp. 82–100, Dec. 2009.
%30
\bibitem{ReidMHT} D. B. Reid, “An Algorithm for Tracking Multiple Targets,” IEEE Transactions on Automatic Control, vol. 24, pp. 843–854, 1979.
%31
\bibitem{MTT_taxonomy} G. W. Pulford, “Taxonomy of multiple target tracking methods,” IEEE Proceedings - Radar, Sonar and Navigation, vol. 152, no. 5, pp. 291+, 2005.
%32
\bibitem{Blackman_MHT} S. S. Blackman, “Multiple hypothesis tracking for multiple target tracking,” IEEE Aerospace and Electronic Systems Magazine, vol. 19, no. 1, pp. 5–18, Jan. 2004.
%33
\bibitem{Leung_GNN} H. Leung, Z. Hu, and M. Blanchette, “Evaluation of multiple radar target trackers in stressful environments,” IEEE Transactions on Aerospace and Electronic Systems, vol. 35, no. 2, pp. 663 –674, apr 1999.
%34
\bibitem{Konstantinova_GNN} P. Konstantinova, A. Udvarev, and T. Semerdjiev, “A study of a target tracking algorithm using global nearest neighbor approach,” in Proceedings of the 4th International Conference on Computer Systems and Technologies: e-Learning, ser. CompSysTech ’03. New York, NY, USA: ACM, 2003, pp. 290–295.
%35
\bibitem{Kuhn1955} H. W. Kuhn, “The hungarian method for the assignment problem,” Naval Research Logistics Quarterly, vol. 2, pp. 83 –97, 1955.
%36
\bibitem{Munkres_rectangular} F. Bourgeois and J.-C. Lassalle, “An extension of the munkres algorithm for the assignment problem to rectangular matrices,” Commun. ACM, vol. 14, no. 12, pp. 802–804, Dec. 1971.
%37
\bibitem{kalman1960} J. B. MacQueen, “Some methods for classification and analysis of multivariate observations,” in Proc. of the 5th Berkeley Symposium on Mathematical Statistics and Probability, L. M. L. Cam and J. Neyman, Eds., vol. 1. University of California Press, 1967, pp. 281–297.
%38
\bibitem{Welch95anintroduction} T. Hastie, R. Tibshirani, and J. Friedman, The Elements of Statistical Learning: Data Mining, Inference, and Prediction, Second Edition, ser. Springer Series in Statistics. Springer, 2009.
%39
\bibitem{tidongle} Texas Instruments. A USB-enabled system-on-chip solution for 2.4 GHz IEEE 802.15.4 and ZigBee applications. [Online]. Available: http://www.ti.com/lit/ds/symlink/cc2531.pdf
%40
\bibitem{802_15_4} IEEE 802.15.4 standard technical specs. [Online]. Available: http://www.ieee802.org/15/pub/TG4Expert.html
%41
\bibitem{srinivasan2006} K. Srinivasan, P. Dutta, A. Tavakoli, and P. Levis, “Understanding the causes of packet delivery success and failure in dense wireless sensor networks,” in Proceedings of the 4th International Conference on Embedded Networked Sensor Systems, 2006, pp. 419–420.
%42
\bibitem{original_OMAT} J. Hoffman and R. Mahler, “Multitarget miss distance via optimal assignment,” Systems, Man and Cybernetics, Part A: Systems and Humans, IEEE Transactions on, vol. 34, no. 3, pp. 327 – 336, May 2004.
%43
\bibitem{OSPA_metric} D. Schuhmacher, B.-T. Vo, and B.-N. Vo, “A consistent metric for performance evaluation of multi-object filters,” Signal Processing, IEEE Transactions on, vol. 56, no. 8, pp. 3447 –3457, aug. 2008.
%44
\bibitem{Zhao_directionality} Y. Zhao, N. Patwari, P. Agrawal, and M. Rabbat, “Directed by directionality: Benefiting from the gain pattern of active rfid badges,” IEEE Transactions on Mobile Computing, vol. 11, no. 5, pp. 865 –877, May 2012.

\end{thebibliography}

%==================================================================================================%
%  Biography
%==================================================================================================%
\begin{biography}[{\includegraphics[width=1.0in,height=1.30in]{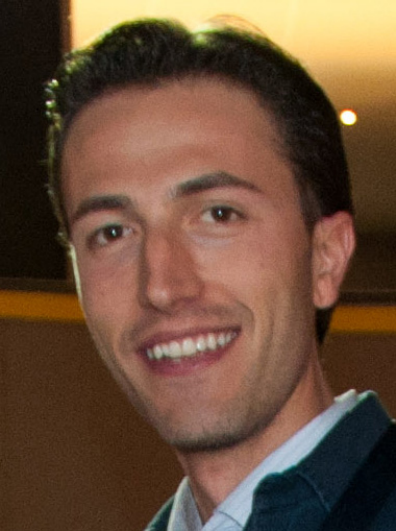}}]{Maurizio Bocca}
received the B.Sc. (2003) and M.Sc. (2006) degrees in Computer Science Engineering from the Politecnico di Milano (Milan, Italy), and the Ph.D. (2011) in Electrical Engineering from Aalto University (Helsinki, Finland). In 2012, he has joined as a post doc the Sensing and Processing Across Networks (SPAN) Lab at the University of Utah (Salt Lake City, Utah, USA), where he is conducting research in the area of RF sensor networks for device-free localization, context awareness and elder care. His research interests include distributed and adaptive algorithms for wireless sensor networks and smart protocols for large-scale deployments of sensor networks in real-world scenarios.
\end{biography}
\begin{biography}[{\includegraphics[width=1.0in,height=1.30in]{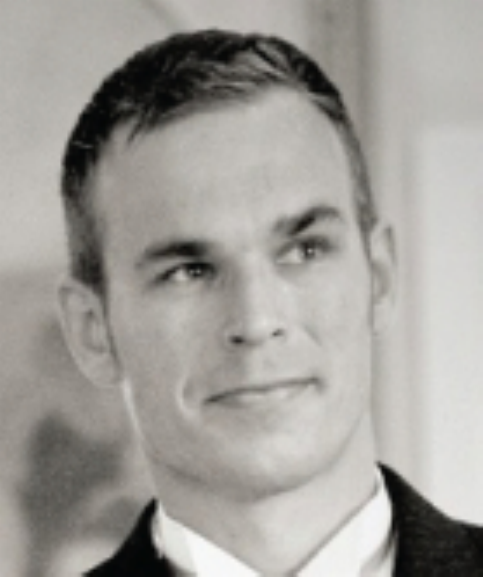}}]{Ossi Kaltiokallio}
received the B.Sc and M.Sc degrees in electrical engineering from Aalto University, School of Electrical Engineering, Helsinki, Finland, both in 2011. He is currently a Ph.D. student with the Department of Automation and Systems Technology, Aalto University School of Electrical Engineering. He is a member of the Wireless Sensor Systems Group at Aalto University. His current research interests include RSS based localization; signal processing, and design and implementation of embedded wireless systems.
\end{biography}
\begin{biography}[{\includegraphics[width=1.0in,height=1.30in]{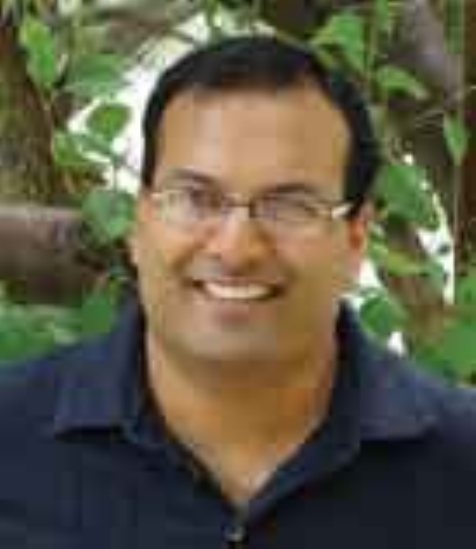}}]{Neal Patwari}
received the B.S. (1997) and M.S. (1999) degrees from Virginia Tech, and the Ph.D. from the University of Michigan, Ann Arbor (2005), all in Electrical Engineering. He was a research engineer in Motorola Labs, Florida, between 1999 and 2001.  Since 2006, he has been at the University of Utah, where he is an Associate Professor in the Department of Electrical and Computer Engineering, with an adjunct appointment in the School of Computing.  He directs the Sensing and Processing Across Networks (SPAN) Lab, which performs research at the intersection of statistical signal processing and wireless networking. Neal is the Director of Research at Xandem, a Salt Lake City-based technology company.  His research interests are in radio channel signal processing, in which radio channel measurements are used to benefit security, networking, and localization applications.  He received the NSF CAREER Award in 2008, the 2009 IEEE Signal Processing Society Best Magazine Paper Award, and the 2011 University of Utah Early Career Teaching Award. He is an associate editor of the IEEE Transactions on Mobile Computing.
\end{biography}
\begin{biography}[{\includegraphics[width=1.0in,height=1.30in]{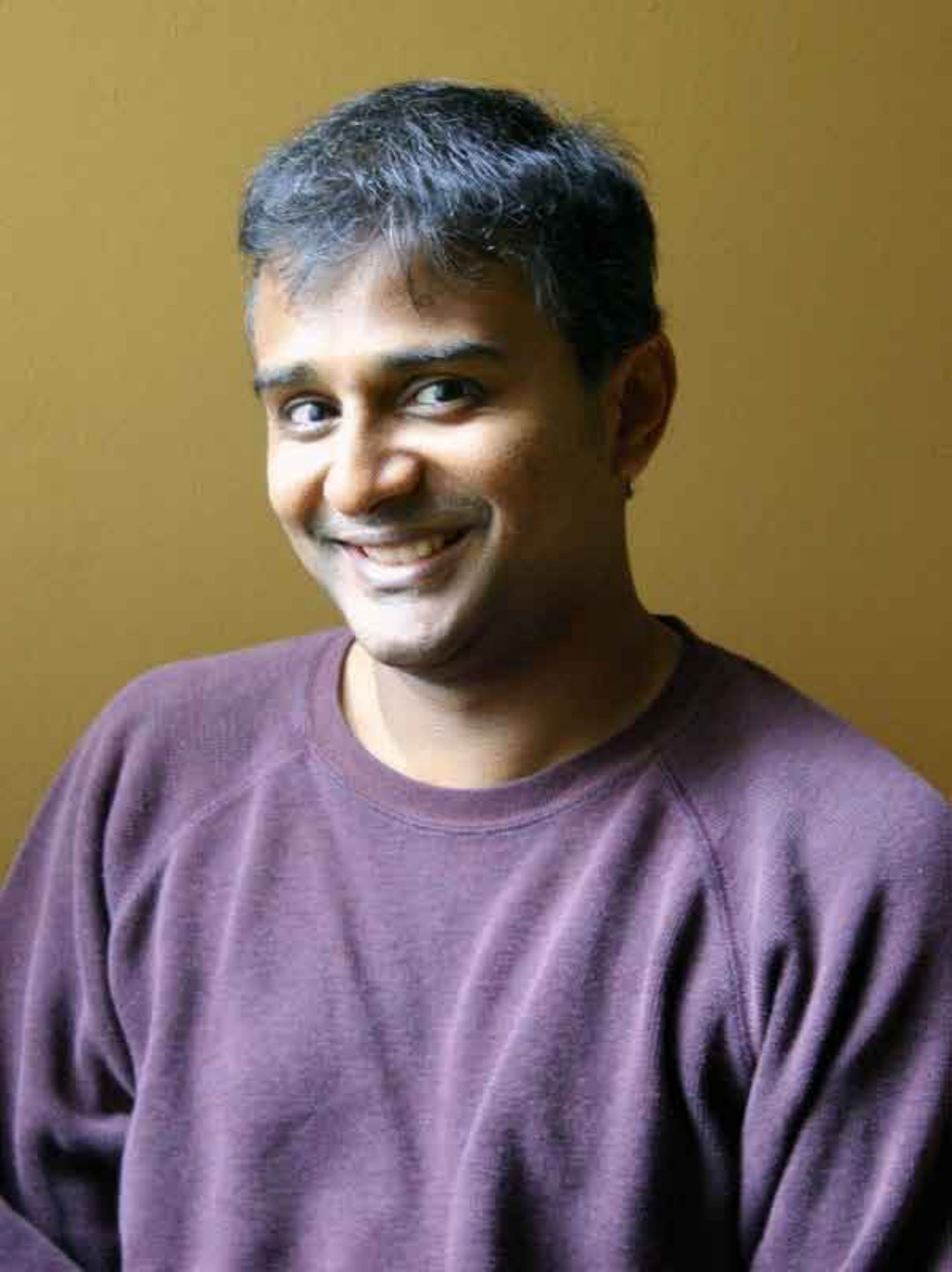}}]{Suresh Venkatasubramanian}
received his Ph.D from Stanford University in 1999. After spending seven years at AT\&T Labs, he came to the University of Utah in 2006, where he is now the Warnock Assistant Professor in the School of Computing. His interests include algorithms, computational geometry, data mining and the challenges of large data processing. He received an NSF CAREER award in 2009.
\end{biography}

    %\epsfig{figure=sensitivity_GNN.eps,width=\columnwidth,height=5.25cm}
    %\includegraphics[width=\columnwidth,height=5.25cm]{sensitivity_GNN.pdf}

% that's all folks
\end{document}